# An Evaluation Framework for Network IDS/IPS Datasets: Leveraging MITRE ATT&CK and Industry Relevance Metrics

Adrita Rahman Tory, and Khondokar Fida Hasan*

*Abstract*—The performance of Machine Learning (ML) and Deep Learning (DL)-based Intrusion Detection and Prevention Systems (IDS/IPS) is critically dependent on the relevance and quality of the datasets used for training and evaluation. However, current AI model evaluation practices for developing IDS/IPS focus predominantly on accuracy metrics, often overlooking whether datasets represent industry-specific threats. To address this gap, we introduce a novel multi-dimensional framework that integrates the MITRE ATT&CK knowledge base for threat intelligence and employs five complementary metrics that together provide a comprehensive assessment of dataset suitability. Methodologically, this framework combines threat intelligence, natural language processing, and quantitative analysis to assess the suitability of datasets for specific industry contexts. Applying this framework to nine publicly available IDS/IPS datasets reveals significant gaps in threat coverage, particularly in the healthcare, energy, and financial sectors. In particular, recent datasets (e.g., CIC-IoMT, CIC-UNSW-NB15) align better with sector-specific threats, whereas others, like CICIoV-24, underperform despite their recency. Our findings provide a standardized, interpretable approach for selecting datasets aligned with sector-specific operational requirements, ultimately enhancing the real-world effectiveness of AI-driven IDS/IPS deployments. The efficiency and practicality of the framework are validated through deployment in a real-world case study, underscoring its capacity to inform dataset selection and enhance the effectiveness of AI-driven IDS/IPS in operational environments.

## I. INTRODUCTION

Intrusion Detection Systems and Intrusion Prevention Systems have become critical components of cybersecurity infrastructures, significantly enhancing an organization's ability to detect and mitigate cyber threats. According to a recent report by Verified Market Research, the global Intrusion Detection and Prevention Systems (IDPS) market was valued at approximately USD 5.8 billion in 2024 and is projected to reach USD 18.6 billion by 2032, exhibiting a compound annual growth rate (CAGR) of 13.2% from 2026 to 2032 (Verified Market Research, 2024). This significant market growth underscores the escalating importance of effective IDS/IPS solutions in mitigating increasingly sophisticated cyber threats across organizational environments.

The integration of Machine Learning (ML) and Deep Learning (DL) into IDS/IPS has considerably improved threat detection rates, enabling faster response times and reducing false positives. Despite significant algorithmic advancements in ML and DL-based IDS/IPS, practical effectiveness remains heavily contingent on dataset quality and contextual relevance (Gharib et al., 2016), (Jindal and Anwar, 2021),(Vahidi et al., 2022), (Verma et al., 2019). Traditional assessment methodologies that rely predominantly on accuracy-based metrics often fail to ensure operational effectiveness, as they neglect whether datasets genuinely represent contemporary, industry-specific threat landscapes (Kasongo and Sun, 2020),(Harahsheh et al., 2024). This oversight is particularly problematic when deploying IDS/IPS solutions in specialized sectors such as healthcare, finance, energy, or retail, where threat vectors vary significantly (Khraisat et al., 2019; Ferriyan et al., 2021; Harahsheh et al., 2024).

Current IDS dataset evaluation frameworks commonly focus on metrics such as detection accuracy, false positive rates, and detection speeds (Gharib et al., 2016; Sharafaldin et al., 2018). However, these frameworks typically do not measure the representativeness of the features of the dataset in relation to real operational contexts. Consequently, even datasets that yield high accuracy in controlled environments may perform inadequately when exposed to real-world threats, undermining operational cybersecurity (Hesford et al., 2024; Kenyon et al., 2020; Song et al., 2020).

The MITRE ATT&CK framework, widely recognized as a comprehensive taxonomy of adversarial behaviors, provides a standardized structure to categorize cyberattacks using tactics, techniques, and procedures (TTPs) (MITRE). Although the use of MITRE ATT&CK for dataset labeling is becoming increasingly common (Bagui et al., 2023), systematic integration into dataset evaluation frameworks remains limited (Husari et al., 2017; Kurniawan et al., 2020; Nari and Ghorbani, 2013). Existing frameworks that evaluate the quality of the dataset, such as those proposed by Gharib et al. (Gharib et al., 2016), Ring et al. (Ring et al., 2019), and Hesford (Hesford et al., 2024), have addressed general dataset characteristics, yet they rarely measure dataset alignment explicitly against industry-specific threats (Li et al., 2021).

To bridge these gaps, this article introduces a comprehensive quantitative evaluation framework tailored to assess IDS/IPS datasets based on five critical metrics. Attack Relevance Score (ARS), Temporal Relevance Score (TRS), Technical Environment Relevance Score (TeRS), Ethical Compliance Score (ECS), and Data Quality Score (DQS). Specifically, our framework integrates MITRE ATT&CK and advanced Natural Language Processing (NLP) techniques to systematically map dataset attack vectors to real-world adversarial techniques, facilitating an objective assessment of industry-specific threat relevance. Through this innovative approach, organizations can more effectively select and leverage datasets that are genuinely representative of their operational threat landscape.

Overall, the primary contribution of this work is twofold. *First,* we propose and develop a novel comprehensive framework for evaluating datasets to be used in developing AI-based IDS/IPS. This framework:

a. Integrates, for the first time, the MITRE ATT&CK framework with semantic similarity techniques to systematically map dataset attacks to industry-specific threat landscapes, thereby enabling objective and operationally aligned dataset selection;

b. Extends beyond technical evaluation by incorporating a comprehensive ethical compliance dimension that assesses privacy protection, consent mechanisms, and regulatory adherence, thus promoting transparency and trustworthiness; and

c. Introduces comprehensive five complementary metrics, ie., Attack Relevance Score, Temporal Relevance Score, Technical Environment Relevance Score, Ethical Compliance Score, and Data Quality Score, to evaluate dataset suitability for real-world deployment rather than laboratory performance alone.

*Second*, we empirically validate the framework in real-world settings. We apply it to nine benchmark IDS/IPS datasets, revealing substantial sector-specific coverage gaps, and further demonstrate its predictive capability through live traffic analysis using Zeek monitoring.

The remainder of the paper is organized as follows: Section II provides a detailed review of the relevant literature and foundational concepts. Section III outlines the methodological details of our proposed framework development. Section IV presents the implementation methodology and industry-specific weighting derivation. Section V provides comprehensive results and analysis of the framework application to benchmark datasets. Section VI discusses the boundaries of applicability and limitations of the proposed approach. Section VII demonstrates empirical validation through a live traffic case study. Finally, Section VIII concludes the paper and highlights future research directions.

## II. BACKGROUND ANALYSIS

Cybersecurity threats have escalated dramatically in recent years, necessitating robust detection and prevention mechanisms. IDS systems are essential components for detecting anomalous activities that may indicate security breaches.

A key challenge in IDS implementation lies in assessing how comprehensively a system can detect real-world attack types within a specific operational domain. This necessitates datasets that not only support high detection performance in laboratory settings but also contain features representative of real threats encountered in domains such as healthcare, finance, or energy (Hesford et al., 2024; Kenyon et al., 2020). The disconnect between experimental success and deployment viability is often due to datasets that are either outdated or lack domain-specific relevance (Song et al., 2020).

The evolution of IDS datasets reflects ongoing efforts to address these limitations. Foundational datasets like Knowledge Discovery and Data Mining Cup 1999 (KDD-CUP'99) (McHugh, 2000; MIT Lincoln Laboratory, 1999) laid the groundwork but were criticized for synthetic traffic and poor representativeness (Ring et al., 2019; Tavallaee et al., 2009). Improvements came with NSL-KDD (a refined version of KDD-CUP'99), which addressed statistical issues by balancing class distributions (Tavallaee et al., 2009; Sapre et al., 2019). More sophisticated datasets emerged with the University of New South Wales Network Behavior 2015 (UNSW-NB15) and the Canadian Institute for Cybersecurity Intrusion Detection System 2017 (CICIDS2017), introducing nuanced attack scenarios and realistic network behaviors (Moustafa and Slay, 2015; Khraisat et al., 2019; Sharafaldin et al., 2018). Recent industry-specific datasets have emerged, including Telemetry of Network Internet of Things (ToN-IoT) (Moustafa, 2021) for IoT environments, Canadian Institute for Cybersecurity Internet of Medical Things (CIC-IoMT) (Dadkhah et al., 2024) for medical devices, and Canadian Institute for Cybersecurity Internet of Things 2022 (CICIoT 2022) (Dadkhah et al., 2022) for general IoT applications, each prioritizing domain-specific threats. However, even specialized datasets demonstrate varying coverage of sector-relevant threats. For instance, comprehensive gap analysis of energy sector datasets has revealed that while some datasets achieve high coverage > 90% when combined, individual datasets like CIC-IDS2017 show coverage as low as 0.38 for critical energy infrastructure threats (Tory et al., 2025). Table I provides a comprehensive comparison of the main IDS datasets for 2009-2024, highlighting their characteristics in these metrics. This analysis reveals both the progress made in the development of the dataset and the remaining gaps in industry-specific representation and the relevance of the features.

TABLE I
Comprehensive IDS Dataset Comparison (2009–2024)

| Criteria Category | NSL-KDD | UNSW-NB15 | CICIDS2017 | CICIDS2018 | UWF-ZeekData22 | CIC-IoMT |
|---|---|---|---|---|---|---|
| **Basic Information** | | | | | | |
| Year of Creation | 2009 | 2015 | 2017 | 2018 | 2022 | 2024 |
| Number of Features | 41 | 49 | 80 | 80 | Several files and several features per file | 39 |
| Number of Attack Types | 4 | 9 | 15 | 7 | 10 | 18 |
| Label Format | Multiclass | Multiclass | Multiclass | Multiclass | Multiclass | Multiclass |
| Attack Types Documented | Yes | Yes | Yes | Yes | Yes | Yes |
| **Data Quality** | | | | | | |
| Feature Relevance | Moderate | High | High | High | High | High |
| Data Consistency | Moderate | High | High | High | High | High |
| Presence of Noise/Outliers | Moderate | Low | Moderate | Moderate | Low | Low |
| Accuracy of Labels | Good | Good | Good | Good | Excellent | Good |
| Preprocessing Effort | High | Medium | Medium | Medium | Medium | Medium |
| **Representativeness** | | | | | | |
| Real-World Traffic Similarity | Limited | Moderate | High | High | High | High |
| Attack Diversity | Moderate | High | High | High | High | High |
| Network Environment Variety | Limited | Moderate | High | High | High | High |
| Temporal Coverage | Limited | Moderate | Comprehensive | Comprehensive | Moderate | Comprehensive |
| Inclusion of IoT Traffic | No | No | No | Partial | No | Yes |
| Scalability to Cloud | No | Limited | Partial | Partial | Limited | Yes |
| **Ethical Considerations** | | | | | | |
| Privacy Preservation | No | Partial | Partial | Partial | Yes | Yes |
| Consent in Data Collection | No | Partial | Partial | Partial | Yes | Yes |
| Potential for Misuse | Medium | Medium | Medium | Medium | Low | Medium |
| Transparency in Creation | Moderate | High | High | High | High | High |
| **Technical Performance** | | | | | | |
| Baseline Accuracy | 97.11% | 99.97% | 99.56% | 99.75% | 99.9% | 98.7% |
| False Positive Rate | 0.62% | 5.29% | 0.12% | 0.52% | 0.08% | 0.3% |
| Detection Rate (Zero-Day) | Low | Moderate | High | High | High | High |
| **Framework Alignment** | | | | | | |
| MITRE ATT&CK Coverage | None | Partial | Partial | Partial | Explicit | Explicit |
| Industry/Domain Focus | General | General | General | Cloud | Cloud | Medical IoT |

Previous evaluation frameworks have attempted to address dataset quality concerns. Gharib et al. (Gharib et al., 2016) proposed 11 criteria specifically for IDS/IPS datasets, while Ring et al. (Ring et al., 2019) identified 15 properties categorized into data volume and recording environments. Hesford (Hesford et al., 2024) emphasized dataset selection criteria including modern threat representation, realism, diversity, and data quality. Despite their contributions, these frameworks rarely incorporate threat modeling or adversarial behavior alignment, limiting their ability to evaluate how well a dataset represents real attack patterns encountered in distinct industry environments. (Li et al., 2021).

The MITRE ATT&CK knowledge base has become the de facto lingua franca for describing adversary TTPs. By assigning every technique a stable identifier and rich, version-controlled narrative, ATT&CK enables researchers to align heterogeneous log sources, threat-intelligence feeds, and IDS/IPS alerts under a single semantic umbrella, thereby facilitating cross-study comparison and reproducibility (MITRE). The popularity of ATT&CK visual "heat-maps" and the widespread inclusion of attack_technique_id metadata in modern rulesets underscore its practical utility for both operational blue-teaming and academic evaluation (Hutchins et al., 2010; National Institute of Standards and Technology, 2006). Recent datasets like UWF-ZeekData22 leverage MITRE ATT&CK for more standardized labeling (Bagui et al., 2023), though systematic integration into dataset evaluation remains limited (Husari et al., 2017; Hasan et al., 2018; Kurniawan et al., 2020; Nari and Ghorbani, 2013).

Although the taxonomy itself is technology-agnostic, empirical studies show that the distribution of techniques invoked by real-world intrusions is highly sector-specific. Credential-stuffing, business-logic abuse, and API fraud dominate incident reports in financial platforms, whereas ransomware operators in healthcare environments more frequently employ lateral movement through DICOM or HL7 services and privilege escalation on medical IoT gateways (National Institute of Standards and Technology, 2018; Sizan et al., 2025; Health Information Trust Alliance, 2020; Federal Financial Institutions Examination Council, 2015). Treating these divergent threat landscapes as homogenous not only dilutes risk prioritisation but can also bias machine-learning evaluation toward unrealistic attack mixes. Recent works therefore advocate weighting ATT&CK techniques by industry relevance, either via manual expert curation (CISA, 2021) or automated mining of campaign reports and CVE descriptions (Daniel et al., 2025; Borisenko et al., 2024).

## III. Framework Development

To ensure a comprehensive industry-specific assessment of IDS/IPS datasets, we identified five key evaluation metrics: ARS, TRS, TeRS, ECS and DQS. These criteria were selected based on their impact on real-world IDS deployment and their ability to quantify dataset suitability beyond traditional accuracy metrics.

The selection of appropriate evaluation criteria for IDS/IPS datasets represents a foundational step in developing a comprehensive assessment framework. Our criteria selection process was guided by three primary considerations: (1) addressing documented limitations in existing evaluation approaches, (2) incorporating industry-specific threat relevance measures, and (3) ensuring holistic assessment across both technical and non-technical metrics. After a thorough analysis of the literature and industry requirements, we identified five key metrics for assessment:

### A. The Attack Relevance Score

The Attack Relevance Score is specifically designed to evaluate the contextual relevance of intrusion detection datasets in the context of industry-specific threat landscapes. This metric aims to fulfill the critical requirement of assessing dataset appropriateness by considering the degree to which the attack vectors represented within the dataset correspond to the threats that organizations in particular industries are likely to encounter.

*1) Methodology:* The foundational principle underlying the ARS metric is that the value of an intrusion detection dataset is significantly determined by its alignment with the real threat landscape pertinent to specific industries. ARS seeks to operationalize this concept by incorporating industry-specific threat intelligence derived from the MITRE ATT&CK framework. By focusing on how well the attack vectors within a dataset align with the sector-specific threats identified through the MITRE ATT&CK framework, the ARS metric offers a more insightful evaluation of a dataset's practical utility for organizations operating in different industries.

Industry-specific threat-intelligence collection constitutes the first methodological phase in our ARS calculation. Because the MITRE ATT&CK Cyber Threat Intelligence (CTI) repository does not explicitly label threat actors by sector, we operationalise the notion of an "industry" through a curated keyword approach. Domain-specific tokens are matched against the name, description, and alias fields of threat actors in the ATT&CK CTI dataset (MITRE). This pragmatic keyword filtering is consistent with the open-source, CTI-driven workflows described in CISA's *Best Practices for MITRE ATT&CK Mapping* (CISA, 2021), which recommends mining threat-intelligence artifacts for behavioral indicators before technique assignment.

Actors that satisfy the keyword filter are treated as *industry-relevant*. We enumerate their associated techniques, compute usage frequencies, and normalize the counts to produce industry-specific weights $\{w_i^{industry}\}$. These weights capture the relative prevalence and criticality of each MITRE technique within the target sector's threat landscape. Figure 2 summarizes the four-step pipeline: keyword definition, threat-actor identification, technique extraction, and frequency-based weight assignment.

The second methodological phase involves mapping each attack type in the dataset to its corresponding MITRE ATT&CK technique. This mapping employs semantic similarity analysis using sentence embeddings to establish connections between attack labels and MITRE techniques. For each attack type in the dataset, contextual information is generated based on statistical properties of its instances. This attack context is then compared against MITRE technique descriptions using cosine similarity measurements. The mapping process incorporates industry weights as a modulating factor, prioritizing techniques that are more relevant to the specified industry when multiple candidates achieve similar semantic similarity scores. The process is illustrated in Figure 3.

Sentence embeddings are fundamental to the ARS

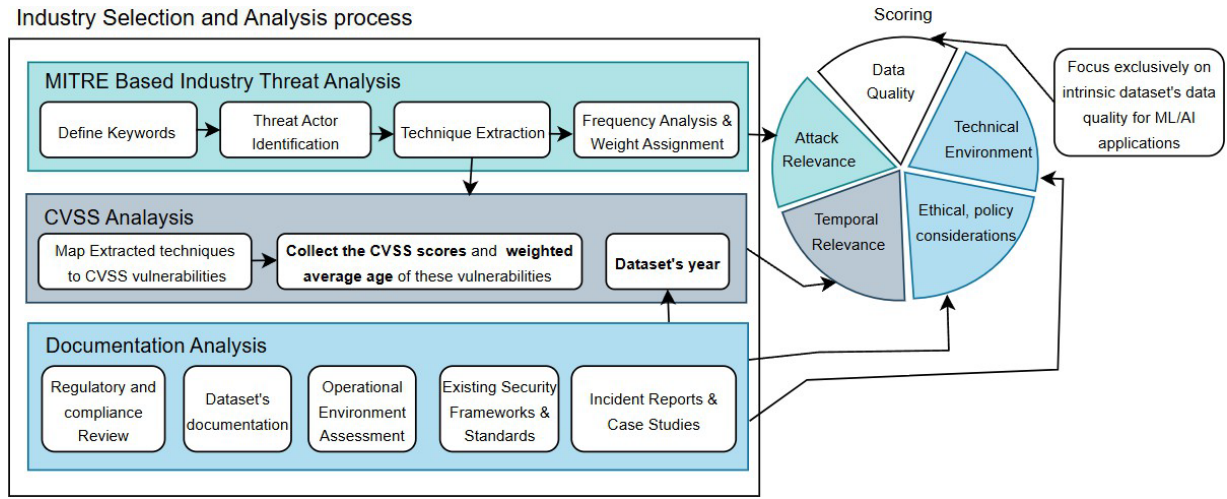

Fig. 1. Overview of the IDS/IPS Dataset Evaluation Framework

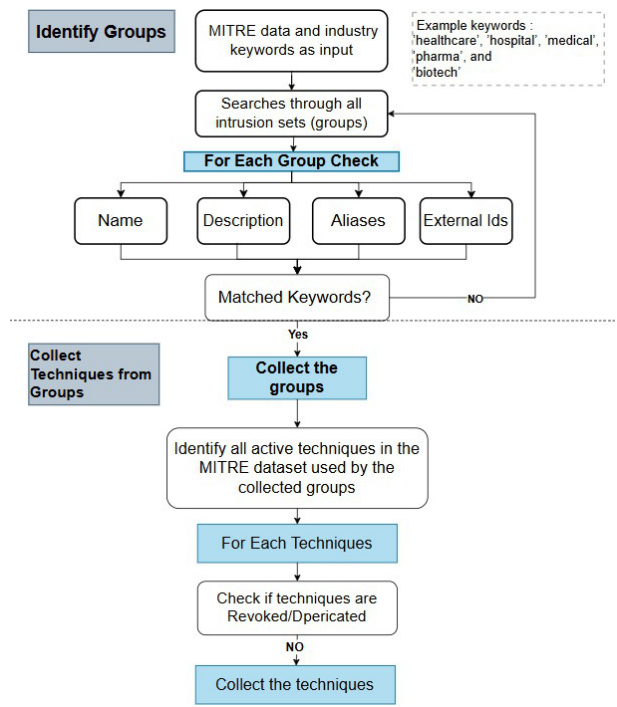

Fig. 2. Extraction of Industry-Specific MITRE ATT&CK Techniques

metric's ability to perform semantic similarity analysis. Cosine similarity serves as the primary metric for measuring the proximity and thus semantic similarity between these sentence embeddings (Mikolov et al., 2013). It calculates the cosine of the angle between two embedding vectors, providing a score that ranges from -1 (completely dissimilar) to 1 (perfectly similar), with 0 indicating no similarity (Mikolov et al., 2013). For sentence embeddings, which are often normalized to unit length, the cosine similarity is equivalent to the dot product of the vectors. This metric is particularly suitable for text comparison as it focuses on the orientation of the vectors, making it less sensitive to differences in the length of the text or the frequency of specific words (Mikolov et al., 2013).

The strategic use of sentence embeddings and cosine similarity allows the ARS metric to perform a more sophisticated and context-aware mapping of attack types to MITRE techniques. This approach transcends the limitations of simple keyword-based matching by considering the underlying semantic meaning of both the attack descriptions in the dataset and the technique descriptions in the MITRE ATT&CK framework.

Seventeen state-of-the-art language models were benchmarked; the compact all-MiniLM-L6-v2 model offered the best trade-off between accuracy(64.3% exact-match on the CICIDS2018 ground truth) and inference cost, and is therefore adopted by default. The framework, however, exposes the model choice as a parameter so that practitioners can substitute a fine-tuned or domain-specific encoder when available. This evaluation used a ground truth dataset comprising 14 attack types from the CICIDS2018 dataset to assess the precision of each model in mapping these attacks to their corresponding MITRE ATT&CK techniques. The results of this comparative analysis indicated that the sentence-transformers/all-MiniLM-L6-v2 model demonstrated the best overall performance for this specific task.

The all-MiniLM-L6-v2 model is a compact and effi-cient transformer-based model from the Sentence Transformers library. It is based on Microsoft's MiniLM architecture, known for its ability to generate high-quality sentence embeddings with a relatively small model size and fast inference speed. The model consists of six transformer layers and produces 384-dimensional embedding vectors for input sentences. Despite its compact design, all-MiniLM-L6-v2 has shown strong performance on a variety of semantic tasks, including semantic similarity, text clustering, and information retrieval. The version 2 iteration of this model has been further optimized through fine-tuning on large datasets like MS MARCO, enhancing its semantic understanding capabilities.

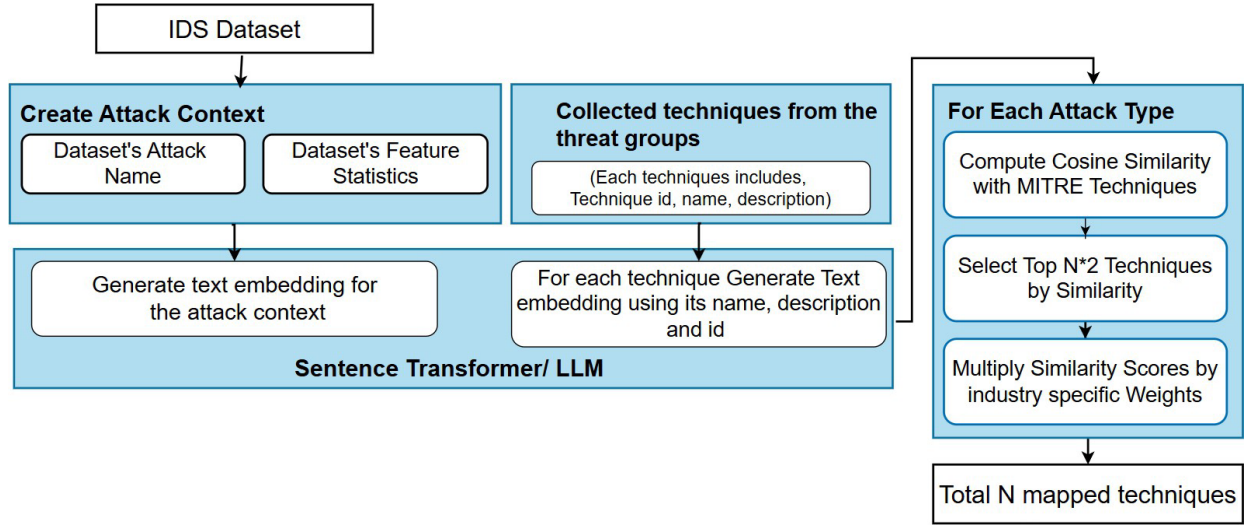

Fig. 3. Semantic Mapping of Dataset Attack Types to MITRE ATT&CK Techniques

The study evaluated multiple embedding models to determine their effectiveness in mapping various cyber-attack techniques from the CICIDS2018 dataset with our generated attack context to their corresponding MITRE ATT&CK techniques. The ground truth mappings, established by Borisenko et al. in their paper (Borisenko et al., 2024), served as the benchmark for assessment. The evaluation methodology measured each model's performance using exact matches (correct identification of the primary technique), partial matches (identification of any correct technique), and overall accuracy.

The core methodology, which maps textual descriptions to MITRE ATT&CK techniques using sentence embeddings and cosine similarity, is fundamentally dataset-agnostic. While this study selected all-MiniLM-L6-v2 as a practical default based on its favorable accuracy-to-cost ratio on CICIDS2018, the pipeline permits seamless substitution. Practitioners can integrate alternative encoders, such as fine-tuned or domain-specific models, by simply updating the model identifier without altering the pipeline. This modular design decouples the model choice from the underlying mapping strategy, providing a reproducible method for encoder selection across diverse datasets.

Results indicated substantial variability in performance across different models, with the acedev003/gte-small-mitre model achieving the highest overall accuracy of 78.57%, followed by sentence-transformers/all-MiniLM-L12-v2 at 71.43%. The basel/ATTACK-BERT and sentence-transformers/all-MiniLM-L6-v2 models both demonstrated moderate effectiveness with identical overall accuracy rates of 64.29%. Notably, contrary to what might be expected, security-specific models like markusbayer/CySecBERT (14.29%), ehsanaghaei/SecureBERT (0%), and jackaduma/SecBERT (0%) performed poorly compared to general-purpose embedding models. The most successful models demonstrated stronger capabilities in correctly identifying denial-of-service attack techniques (T1499, T1499.002) compared to brute force techniques (T1110, T1110.001). The results are summarized in Table II.

*2) Formula and Calculation:* The semantic similarity mapping process generates technique mappings t(a) for each attack instance through cosine similarity computation in the embedding space. These mappings, combined with empirical frequency data f(a), feed into our ARS calculation to produce industry-contextualized threat scores that reflect both attack prevalence and sector-specific relevance. With these mappings established, for each attack label a, we obtain its mapped technique t(a) by the embedding-based cosine similarity procedure (Algorithm 2) using threshold $\theta_{sim}$. Ties are resolved by the larger industry weight $w_t$. The label frequency $f(a)$ is computed over records excluding the *normal* class.

TABLE II
PERFORMANCE EVALUATION OF EMBEDDING MODELS

| Model Name | Exact Match % | Partial Match % | Overall Accuracy % |
|---|---|---|---|
| acedev003/gte-small-mitre | 21.43 | 57.14 | 78.57 |
| sentence-transformers/all-MiniLM-L12-v2 | 28.57 | 42.86 | 71.43 |
| basel/ATTACK-BERT | 42.86 | 21.43 | 64.29 |
| sentence-transformers/all-MiniLM-L6-v2 | 21.43 | 42.86 | 64.29 |
| all-mpnet-base-v2 | 35.71 | 21.43 | 57.14 |
| multi-qa-mpnet-base-dot-v1 | 7.14 | 50.00 | 57.14 |
| sentence-transformers/multi-qa-mpnet-base-dot-v1 | 0.00 | 50.00 | 50.00 |
| BAAI/bge-m3 | 42.86 | 7.14 | 50.00 |
| Sentence-t5-base | 0.00 | 28.57 | 28.57 |
| paraphrase-albert-small-v2 | 0.00 | 28.57 | 28.57 |
| google/flan-t5-large | 0.00 | 21.43 | 21.43 |
| markusbayer/CySecBERT | 0.00 | 14.29 | 14.29 |
| microsoft/deberta-v3-base | 0.00 | 0.00 | 0.00 |
| ehsanaghaei/SecureBERT | 0.00 | 0.00 | 0.00 |
| jackaduma/SecBERT | 0.00 | 0.00 | 0.00 |
| sarahwei/MITRE-v15-tactic-bert-case-based | 0.00 | 0.00 | 0.00 |
| bencyc1129/mitre-bert-base-cased | 0.00 | 0.00 | 0.00 |

$$\text{ARS}(D) = \frac{\sum_{i=1}^{n_{attack}} f(a_i) \cdot W_{t(a_i)}}{\sum_{i=1}^{n_{attack}} f(a_i) \cdot W_{max}} \tag{1}$$

Where:

· $n_{attack}$ is the number of attack types present in dataset $D$, excluding normal traffic.
· $a_i$ is the $i$-th attack type in the dataset.
· $f(a_i)$ is the relative frequency of attack type $a_i$ in the dataset (i.e., the proportion of rows labeled as $a_i$).
· $t(a_i)$ is the top-matching ATT&CK technique associated with attack $a_i$, derived from the embedding-based semantic similarity mapping process.

· $W_{t(a_i)}$ is the industry-specific weight for technique $t(a_i)$. For unmapped labels where no technique clears threshold $\theta_{sim}$, we assign a conservative fallback weight of $0.10$ to ensure comprehensive threat coverage. This design choice prevents the mathematical elimination (multiplication by zero) of CTI-identified techniques that may be strategically relevant but infrequently observed in training data, prioritizing security completeness over statistical precision while still penalizing unmapped labels; we fix this value for reproducibility.
· $W_{max}$ is the highest industry weight among all techniques in the mapping, serving as the normalization factor.

Because $W_·$ are normalized with $\max_t W_t = 1$, the denominator equals $\sum_{a \in A_D} f(a)$; thus ARS represents the frequency-weighted mean of the sector weights assigned to the mapped techniques. This formulation ensures that datasets with attack vectors highly relevant to the specified industry's threat landscape receive higher ARS scores, indicating optimal alignment with real-world threats faced by organizations in that sector.

Although the computational function returns the result as a percentage, ARS values are consistently reported in their normalized $[0,1]$ form in this paper to maintain alignment with other evaluation metrics. The resulting ARS metric offers a nuanced interpretation framework where scores approaching 1 indicate datasets containing attack vectors highly relevant to the specified industry's threat landscape, suggesting optimal alignment with real-world threats, while scores approaching 0 reveal datasets populated primarily with attack types having minimal relevance to industry-specific threats.

*3) Interpretation:* The resulting ARS metric offers a nuanced interpretation framework. Scores approaching 1 indicate datasets containing attack vectors highly relevant to the specified industry's threat landscape, suggesting optimal alignment with real-world threats faced by organizations in that sector. In contrast, scores approaching 0 reveal datasets populated primarily with attack types having minimal relevance to industry-specific threats, indicating limited practical utility for security applications within that context.

*B. Temporal Relevance Score*

Empirical evidence from a five-year network traffic analysis indicates that traditional machine learning-based intrusion detection models demonstrate limited temporal validity, with effectiveness declining significantly within approximately two weeks post-training (Viegas et al., 2020). Moreover, the research highlights that the datasets predominant in the 2010 research (DARPA98 and KDD-CUP'99) were already a decade old at that time (de Carvalho Bertoli et al., 2022), suggesting a substantial temporal disconnect between the research materials and contemporary threat landscapes. Further investigation utilizing a four-year real network traffic dataset confirms that existing intrusion detection methodologies inadequately address the evolutionary nature of network traffic patterns (dos Santos et al., 2022). These findings collectively indicate that IDS datasets possess restricted periods of relevance due to cybersecurity's inherently dynamic characteristics.

*1) Methodology:* The practical lifetime of an IDS dataset is not predetermined but is contingent on multiple interrelated factors. The emergence rate of new cyber threats and the velocity of evolution of existing attack vectors substantially influence the obsolescence of the data set (Goldschmidt and Chuda´, 2025). Similarly, the representational accuracy of a dataset of contemporary network traffic patterns and protocols significantly impacts its continued utility (Goldschmidt and Chuda´, 2025). The specific application context or research objective further modulates acceptable temporal decay thresholds. Consequently, dataset lifespan represents a variable parameter determined by threat environment evolution and intrinsic dataset attributes. Narrowly focused datasets may experience accelerated irrelevance if their targeted attack vectors diminish in prevalence or are superseded by emerging methodologies.

To quantitatively capture this temporal dimension, we introduce the TRS, employing a logistic-like decay function rather than simple exponential decay. This methodological choice reflects our understanding that while older datasets and techniques inevitably become less relevant over time, they do not instantaneously lose all utility. The TRS calculation integrates two key temporal factors: (1) the chronological age of the data set itself and (2) the temporal currency of the attack techniques represented within the data set, as evidenced by the associated vulnerability reports. This dual temporal consideration provides a nuanced measure that better reflects real-world security operations.

The logistic decay function $\psi(\Delta t)$ is defined as

$$\psi(\Delta t) = \frac{1}{1 + \Delta t/\lambda}, \qquad (2)$$

where $\Delta t$ represents the time difference in years and $\lambda$ is a decay constant that determines the rate at which relevance diminishes over time.

Our methodology employs the Common Vulnerabilities and Exposures (CVE) repository as the primary temporal reference for attack techniques. This selection is justified based on several critical attributes of the CVE program that make it uniquely suitable for temporal analysis in the cybersecurity domain. First and foremost, CVE records provide standardized and verifiable chronological markers for vulnerability discovery and disclosure. Unlike other potential sources of temporal information, such as less formally structured threat intelligence reports or malware timestamps, CVE entries undergo a rigorous validation process before publication, ensuring their temporal accuracy. According to the official MITRE CVE documentation, each entry undergoes a comprehensive review by a CVE Numbering Authority (CNA), which enforces standardized temporal recording practices (CVE Program, 2024).

The CVE program has maintained consistent documentation standards since its inception in 1999, providing over two decades of continuous vulnerability data. This longitudinal consistency allows for reliable temporal analysis across different eras of cybersecurity evolution. Despite its age, the CVE repository offers unparalleled historical depth while maintaining contemporary relevance through ongoing curation. The comprehensive historical data within the CVE repository allows for a thorough understanding of the evolution of vulnerabilities and their relation to attack techniques over a significant timeframe, a feature that is crucial for contextualizing temporal relevance (Mann and Christy, 1999).

Additionally, CVE entries contain rich contextual information about vulnerability types, affected systems, and exploitation methods. This detailed information enables semantic alignment with MITRE ATT&CK techniques, creating a crucial bridge between abstract attack techniques and concrete vulnerability instances. According to Abdeen et al. (2023) (Abdeen et al., 2023), the descriptor fields within CVE records, particularly the description and references sections, provide valuable context that facilitates mapping to specific attack techniques within the MITRE ATT&CK framework.

To establish connections between dataset attack vectors and CVE entries, we employ the SMET (Semantic Mapping of CVE to ATT&CK) model developed by Abdeen et al. (Abdeen et al., 2023). The integration of this model into our framework represents a strategic choice for enhancing the relevance evaluation within the TRS calculation. SMET is specifically designed to automatically map CVE entries to techniques within the MITRE ATT&CK framework based on the textual similarity of their descriptions. This mapping is crucial for understanding the potential impact of vulnerabilities from an attacker's perspective and for assessing the relevance of attack techniques within IDS datasets in the context of known adversarial behaviors (Center for Threat-Informed Defense, 2024).

A significant advantage of SMET is its ability to perform unsupervised mapping of CVE descriptions to MITRE ATT&CK techniques. Unlike other mapping models that rely on manually annotated datasets, SMET leverages the inherent semantic relationships between vulnerability descriptions and attack techniques to establish connections. This unsupervised approach makes SMET highly adaptable to the dynamic nature of cybersecurity threats, as it can learn from new data and evolving language without requiring constant manual updates to training datasets (Abdeen et al., 2023).

*2) Formula and Calculation:* Our TRS calculation process comprises a series of well-defined steps designed to quantify the temporal relevance of an IDS dataset:

We calculate the dataset's chronological age in years relative to the current date:

$$\Delta t_{\text{dataset}} = \frac{\text{current date} - \text{dataset creation date}}{365.25}. \quad (3)$$

We then apply the logistic decay function to determine its baseline temporal relevance:

$$\phi_d = \psi(\Delta t_{\text{dataset}}) = \frac{1}{1 + \Delta t_{\text{dataset}}/\lambda}, \quad (4)$$

where $\lambda$ represents our empirically justified decay constant of 2.0.

For each attack technique identified in the dataset (through our ARS methodology), we use SMET to identify associated CVE entries based on semantic similarity. This step results in a set of CVEs for each technique, with their respective publication dates and CVSS scores (when available).

For each technique with mapped CVEs, a weighted average age is calculated based on the publication dates of the associated vulnerabilities. CVSS scores serve as weights, giving greater emphasis to more severe vulnerabilities:

$$\Delta t_{\text{tech}} = \frac{\sum (\text{CVE age}_i \times \text{CVSS score}_i)}{\sum \text{CVSS score}_i}, \quad (5)$$

This weighted average prioritizes the recency of techniques associated with high-severity vulnerabilities, reflecting their potentially greater impact on the current threat landscape. More critical vulnerabilities are likely to be exploited more actively and thus contribute more significantly to the temporal relevance of the associated attack technique.

The same logistic decay function is applied to each technique's age to determine its individual temporal relevance coefficient:

$$\rho_i = \psi(\Delta t_{\text{tech}}) = \frac{1}{1 + \Delta t_{\text{tech}}/\lambda}, \quad (6)$$

This consistent application of the decay function across both dataset and technique levels ensures methodological coherence throughout the temporal relevance assessment. For techniques without mappable CVE associations or with insufficient temporal data, our framework assigns a default middle relevance coefficient $\rho_i = 0.5$, representing a neutral position that neither unduly penalizes nor inappropriately advantages techniques lacking explicit vulnerability linkages. This methodological choice acknowledges that absence of CVE correlation may result from various factors beyond temporal irrelevance, including novel techniques that have yet to be formally documented within vulnerability repositories.

The comprehensive TRS is computed as a weighted sum of the technique relevance coefficients, modulated by the dataset decay coefficient:

$$\text{TRS}_{\text{industry}} = \phi_d \times \frac{\sum_{i=1}^{n_{\text{tech}}} W_i^{(\text{industry})} \rho_i}{\sum_{i=1}^{n_{\text{tech}}} W_i^{(\text{industry})}}, \quad (7)$$

where

· $W_i^{(\text{industry})}$ is the industry-specific weight for technique $i$, derived from the ARS methodology.
· $\rho_i \in [0, 1]$ is the relevance coefficient of technique $i$ based on its temporal validity.
· $n_{\text{tech}}$ is the total number of techniques considered.
· $\phi_d$ is the dataset decay coefficient that adjusts the score to reflect dataset currency.

This produces a normalized score between 0 and 1, where higher values indicate that the dataset better represents the temporally relevant adversarial behaviors within the target industry.

*3) Parameter Selection and Justification:* The selection of the decay constant $\lambda = 2.0$ in our TRS function was grounded in three key areas of analysis:

**Vulnerability Lifecycle Data:** Empirical research suggests that the half-life of vulnerability relevance in operational security contexts is approximately 2 years (Abdeen et al., 2023) (Kandek, 2009). While some studies found critical vulnerability half-lives as short as 30 days (Abdeen et al., 2023), our approach considers a broader operational security context. The data indicates that the practical relevance of vulnerabilities and associated attack techniques tends to decrease significantly within a two-year timeframe.

**Industry Security Practice Guidelines:** Most industry guidelines recommend reassessing security controls and threat models every 18-24 months (Practical DevSecOps, 2024) (Carbide Security, 2024). This consensus suggests that a two-year window represents a significant period of change in the cybersecurity landscape, supporting our decay constant selection.

**MITRE ATT&CK Framework Evolution:** Analysis of the MITRE ATT&CK framework reveals substantial changes in attacker tactics and techniques over 2-3 year periods (Al-Sada et al., 2023). With framework updates occurring approximately every 6 months, the threat landscape demonstrates significant evolution within this timeframe.

*4) Interpretation:* This formulation ensures that the final temporal relevance assessment integrates both the macro-level chronological position of the dataset (captured by $\phi_d$) and the micro-level currency of its constituent attack techniques (captured by $\rho_i$), while simultaneously accommodating industry-specific threat priorities (captured by $W_i^{\text{(industry)}}$). The resulting score, bounded between 0 and 1, provides security practitioners with a mathematically rigorous yet practically interpretable metric for evaluating dataset temporal relevance within specific operational contexts.

## C. Technical Environment Relevance Score

Technical Environment Relevance Score is designed to assess how well a dataset reflects the technical characteristics of the target deployment environment, ensuring that IDS/IPS systems are tested and trained in a manner that mirrors real operating conditions. By incorporating protocol diversity, traffic complexity, and network environment representation, TeRS evaluates the real-world applicability of datasets, bridging the gap between laboratory performance and operational effectiveness.

*1) Methodology and Criteria Selection:* The criteria selected for TeRS are grounded in research that emphasizes the importance of real-world network environments for the evaluation of cybersecurity systems. Many scholars have acknowledged that the success of IDS/IPS systems depends not only on the quality of the data but also on how well that data represents the environments in which these systems will be deployed.

**Network and Protocol Comprehensiveness:** Research by Ring et al. (Ring et al., 2019) and Gharib et al. (Gharib et al., 2016) stresses the importance of protocol diversity for effective IDS training. IDS systems need to be capable of detecting attacks across a variety of protocols (e.g., HTTP, DNS, FTP) since they function in diverse network environments. As Gharib et al. (Gharib et al., 2016) explain, datasets lacking protocol diversity can lead to an overfitting of models, where the IDS performs well on certain protocols but fails to detect attacks on less common ones. Thus, a diverse protocol representation, as reflected in the Protocol Diversity subcriterion of TeRS, ensures that the dataset prepares IDS models for a broad spectrum of potential attack vectors.

**Data Capture Methodology:** The methodology behind data capture plays a vital role in ensuring the dataset's realism. According to Sharafaldin et al. (Sharafaldin et al., 2018), datasets collected from real network traffic provide a more accurate simulation of network behavior, ensuring the IDS systems trained on them will encounter real-world anomalies. Furthermore, Ring et al. (Ring et al., 2019) note that datasets with synthetic or emulated traffic may not represent the full complexity of live network environments. Therefore, the Diversity of Capture Methods and Granularity and Temporal Coverage in TeRS reflect the need for realistic, long-duration data captures that mirror continuous, real-world network activity.

**Payload and Metadata Analysis:** The depth of payload analysis has been highlighted by several studies, including those by Sharafaldin et al. (Sharafaldin et al., 2018), which argue that the granularity of metadata and payload features is crucial for detecting advanced persistent threats (APTs) and sophisticated attack strategies. Datasets that provide rich payload information enable IDS/IPS systems to go beyond surface-level detection and identify deeper, more subtle forms of attack. Thus, the Payload Detail and Feature Extraction Potential subcriteria in TeRS emphasize the need for detailed, feature-rich datasets to allow for the development of more precise IDS models.

**Recording Environment Characterization:** Sharafaldin et al. (Sharafaldin et al., 2018) and Gharib et al. (Gharib et al., 2016) have also argued that realistic network environments are essential for evaluating IDS/IPS systems. They state that enterprise networks, cloud infrastructures, and IoT environments pose unique challenges for intrusion detection, each with distinct network configurations and attack vectors. Datasets should include traffic patterns and devices that represent the specific types of networks IDS/IPS will be deployed in. The Environment Type and Traffic Authenticity sub-criteria within TeRS address this issue, ensuring that datasets encompass a wide variety of environments.

**Specialized Domain Relevance:** Industry-specific datasets are becoming increasingly important as cybersecurity needs evolve. Research in the field, as highlighted in studies like Goldschmidt and Chudá (Goldschmidt and Chudá, 2025), emphasizes the importance of datasets tailored to sectors like healthcare, finance, and IoT, which have unique protocols and security requirements. The inclusion of industry-specific protocols and specialized network configurations in TeRS ensures that datasets reflect the precise needs of these domains, enabling more accurate testing and training of IDS/IPS systems in these specialized contexts.

Table III presents a comprehensive overview of the evaluation criteria for the Technical Environment Relevance Score, including specific metrics and evaluation scales for each dimension.

*2) Formula and Calculation:* The TeRS is calculated using the following formula:

$$\text{TeRS}_{\text{industry}} = \frac{\sum_{i=1}^{n_{\text{TeRS}}} \text{Score}_i \cdot W_i^{\text{(industry)}}}{\max(\text{Score}) \times \sum_{i=1}^{n_{\text{TeRS}}} W_i^{\text{(industry)}}}, \quad (8)$$

TABLE III
CRITERIA OF TECHNICAL ENVIRONMENT RELEVANCE SCORE

| Criteria | Sub-Criteria | Description | Evaluation Metrics | Evaluation Scale |
|---|---|---|---|---|
| 1. Network and Protocol Comprehensiveness | 1.1 Protocol Diversity | Assesses the variety of network protocols (e.g., TCP/IP, HTTP, DNS) included in the dataset. | Count of unique protocols in dataset | 0-5 (0 = No protocols, 5 = Multiple protocols from various layers) |
| | 1.2 OSI Layer Representation | Evaluates how well the dataset covers all OSI layers, from physical to application. | Percentage of OSI layers covered | 0-5 (0 = No layers, 5 = All layers represented) |
| | 1.3 Inter-protocol Complexity | Measures how well the dataset captures interactions between multiple protocols. | Number of inter-protocol interactions | 0-5 (0 = No interactions, 5 = Complex interactions between multiple protocols) |
| 2. Data Capture Methodology | 2.1 Diversity of Capture Methods | Evaluates the variety of methods used to capture traffic (e.g., live traffic, emulated traffic). | Variety of capture methods (real, emulated, synthetic) | 0-5 (0 = One capture method, 5 = Multiple diverse capture methods) |
| | 2.2 Granularity and Temporal Coverage | Assesses the temporal and spatial coverage of the dataset, ensuring it includes long-duration traffic. | Duration and time intervals of traffic coverage | 0-5 (0 = Short duration, 5 = Long duration and diverse time intervals) |
| | 2.3 Capture Environment Authenticity | Measures the authenticity of the data capture environment (real-world vs. synthetic). | Authenticity of data capture environment | 0-5 (0 = Synthetic data only, 5 = Real-world data capture environment) |
| 3. Payload and Metadata Analysis | 3.1 Payload Detail | Evaluates the depth and richness of payload information captured in the dataset. | Amount of payload data provided | 0-5 (0 = No payload data, 5 = Detailed payload with full data characteristics) |
| | 3.2 Metadata Variety | Assesses the diversity and richness of metadata captured in the dataset, including attributes like IP, ports, timestamps. | Number of metadata attributes captured | 0-5 (0 = No metadata, 5 = Comprehensive metadata including timestamps, IPs, ports) |
| | 3.3 Feature Extraction Potential | Measures the dataset's capacity for advanced feature extraction to aid in IDS/IPS model training. | Range of features available for extraction | 0-5 (0 = No features, 5 = Large range of features for detailed analysis) |
| 4. Recording Environment Characterization | 4.1 Environment Type | Assesses the variety of network environments represented in the dataset (e.g., enterprise, cloud, IoT). | Number of environments covered | 0-5 (0 = One environment, 5 = Multiple diverse environments like IoT, cloud, enterprise) |
| | 4.2 Traffic Authenticity | Evaluates the realism of traffic patterns in the dataset, ensuring that it reflects real-world traffic behavior. | Degree of authenticity in traffic patterns | 0-5 (0 = Synthetic traffic, 5 = Realistic traffic patterns that mimic real-world scenarios) |
| 5. Specialized Domain Relevance | 5.1 Industry-Specific Protocols | Measures how well the dataset reflects protocols specific to the targeted industry (e.g., medical IoT, SCADA). | Inclusion of industry-specific protocols | 0-5 (0 = No industry-specific protocols, 5 = Full inclusion of protocols specific to the industry) |
| | 5.2 Specialized Network Configurations | Assesses whether the dataset includes specialized configurations such as virtualized environments or cloud-native architectures. | Presence of specialized network configs | 0-5 (0 = No specialized configurations, 5 = Includes specific network configurations like cloud, edge, or SCADA) |
| | 5.3 Security Controls Representation | Evaluates how well the dataset represents security controls commonly used in specific industries (e.g., firewalls, encryption). | Coverage of security controls in the dataset | 0-5 (0 = No security controls, 5 = Full representation of industry-standard security mechanisms) |

where $\text{Score}_i \in [0, 5]$ is the score for criterion $i$, $W_i^{(\text{industry})}$ is the industry-specific weight for that criterion, $n_{\text{TeRS}}$ is the number of TeRS criteria, and max_score = 5.

This produces a normalized score between 0 and 1, with higher values indicating greater relevance to the technical environment of the target industry.

*3) Scoring System and Weighting:* The TeRS scoring system follows a structured, weighted approach, where each sub-criterion is evaluated on a scale from 0 to 5, with higher scores representing a greater level of relevance to the target technical environment. To enhance contextual relevance, we incorporate industry-specific weighting factors for each criterion. These weights reflect the varying importance of different technical aspects across sectors such as finance, government, healthcare, and critical infrastructure. For instance, financial institutions may place greater emphasis on payload detail and security controls representation, while government agencies might prioritize specialized network configurations and traffic authenticity.

*4) Interpretation:* The use of this industry-calibrated weighted scoring system allows for a nuanced evaluation, ensuring that the most critical aspects for a specific sector receive appropriate importance in the overall assessment. This approach reflects the real-world trade-offs faced by cybersecurity practitioners who must prioritize certain factors depending on the network they are securing and the industry they operate within.

Additionally, the evaluation scale from 0 to 5 enables clear differentiation between datasets, offering a quantitative measure of their relevance. A dataset that scores high on the industry-weighted TeRS demonstrates a strong alignment with real-world deployment contexts for that specific sector, while a low score indicates that the dataset may be better suited for other industries or less complex environments. This scoring system is intuitive and directly applicable to the decision-making processes of cybersecurity professionals who need reliable, industry-specific datasets to train their IDS/IPS systems for their particular operational contexts.

### D. Ethical Compliance Score

Network traffic data, which forms the core of most IDS datasets, inherently carries a multitude of privacy risks due to the sensitive nature of the information it contains. The metadata associated with network packets, such as source and destination IP addresses, port numbers, and communication protocols, can often be directly linked to specific individuals or organizations, potentially revealing their identities and online activities. Furthermore, the payload of network packets, which contains the real data being transmitted, might include highly sensitive communication content, such as emails, chat messages, browsing history, and personal files. Even seemingly innocuous information, such as traffic flow patterns, the timing of communications, and the frequency of interactions with specific services, can be analyzed to infer user behavior, habits, and even sensitive personal details (Chen et al., 2023).

Fairness is an essential ethical consideration, requiring that datasets be representative of the population or network environments they are intended to model and that they avoid biases that could lead to unfair or discriminatory outcomes in the development or evaluation of IDS. Transparency is also paramount, obligating researchers to be open and honest about how datasets were collected, processed, and anonymized, including any limitations, potential biases, or known privacy risks associated with the data. This transparency allows other researchers to critically evaluate the dataset's suitability for their own work and to understand the context in which any findings based on the dataset were generated (European Union, 2018).

The sharing of datasets, even those that have been anonymized, carries inherent ethical implications due to the potential for misuse or unintended privacy breaches. It is crucial to establish clear data usage agreements and ethical guidelines that govern how shared datasets can be accessed, used, and further disseminated to ensure responsible data handling and prevent potential harm (Andrews et al., 2023) (European Union, 2018).

*1) Methodology and Criteria Selection:* The criteria selected for ECS are grounded in research that emphasizes the importance of ethical considerations in cybersecurity dataset development and usage. These criteria address the complex ethical challenges associated with collecting, processing, and sharing network traffic data for security research.

**Network Data Anonymization:** This dimension addresses the techniques used to protect the privacy of individuals and organizations whose network traffic is included in the dataset. Effective anonymization is paramount to mitigate the risks of re-identification and privacy breaches (Gadotti et al., 2024). The identification and removal of Personally Identifiable Information (PII) such as names, addresses, and social security numbers are critical, with techniques like masking, tokenization, and redaction commonly employed to protect sensitive data (McCallister, 2010). IP address obfuscation is particularly important since addresses can often be linked to individuals or organizations; methods such as prefix-preserving anonymization, randomization, and truncation help mitigate this risk while maintaining the data's utility for analysis. Network packet payloads represent another privacy concern, as they may contain highly sensitive information; sanitization techniques aim to remove or alter this content while preserving the data's security analysis value, though balancing privacy with the need for realistic attack data presents ongoing challenges. Beyond payloads, metadata such as timestamps, port numbers, and protocols can also pose privacy risks if not properly anonymized; approaches like generalization and suppression are typically applied to protect this information while maintaining analytical value.

**Forensic Integrity Preservation:** This dimension focuses on maintaining the trustworthiness and reliability of the dataset for forensic analysis, even after anonymization. Evidence authenticity is crucial, ensuring that data has not been tampered with and that its origin can be verified for forensic investigations; techniques like cryptographic hashing provide mechanisms to verify data integrity throughout the dataset lifecycle (Akbarfam et al., 2023). Implementing data tampering prevention measures, including access controls and write protection mechanisms, is essential to maintain the dataset's forensic value by preventing unauthorized modification or deletion. Documentation of the chain of custody, tracking the handling and storage of the dataset from collection through analysis, is vital for legal admissibility and integrity assurance; maintaining a clear and unbroken chain of custody is considered a best practice in forensic data management. Verifiable data provenance further enhances trustworthiness by tracking the origin and history of the data, including any transformations or anonymization steps applied; sophisticated data provenance solutions and metadata management systems can provide the necessary tracking capabilities for complex forensic applications (Kenyon et al., 2020).

**Consent and Notification Mechanisms:** This dimension addresses the ethical considerations related to obtaining consent for data collection and informing individuals about network monitoring. Organizational consent represents a fundamental ethical requirement, involving obtaining permission from relevant organizations or network owners before collecting and sharing network traffic data; comprehensive data sharing agreements should clearly outline the terms of use and responsibilities of all parties. Network monitoring disclosure ensures transparency about monitoring practices through privacy policies that clearly articulate what data is being collected, how it will be used, and with whom it might be shared. Usage limitation principles dictate that data should only be used for the specific purposes for which it was collected and with appropriate consent; repurposing data without explicit permission raises significant ethical concerns that could undermine trust in research. Opt-out provisions respect individual autonomy and privacy rights by providing mechanisms to decline participation in data collection or certain types of processing; clear and accessible opt-out mechanisms should be available to all affected parties (de Man et al., 2023) (Alter and Gonzalez, 2018).

**Ethical Research Use:** This dimension focuses on the ethical principles that should guide the use of IDS datasets in research. Research purpose limitations ensure datasets are used solely for legitimate research purposes, such as improving intrusion detection techniques or understanding network security threats; misuse of data for unintended purposes is considered unethical and potentially harmful. Academic integrity requirements compel researchers to adhere to the highest standards of honesty, including proper citation, avoiding plagiarism, and accurately reporting findings; the integrity of the research process remains paramount for maintaining trust in security research. Security research guidelines provide frameworks for responsible research practices, including vulnerability disclosure protocols and harm minimization strategies (Alter and Gonzalez, 2018).

Table IV provides a comprehensive overview of the evaluation criteria for privacy and ethical considerations in IDS/IPS datasets, including specific metrics and evaluation scales for each dimension.

*2) Formula and Calculation:* The ECS is calculated using industry-specific weights that reflect varying ethical priorities across sectors:

$$\text{ECS}_{\text{industry}} = \frac{\sum_{i=1}^{n_{\text{ECS}}} \text{Score}(c_i) \, W_i^{(\text{industry})}}{\max(\text{Score}) \times \sum_{i=1}^{n_{\text{ECS}}} W_i^{(\text{industry})}}, \quad (9)$$

where $\text{Score}(c_i) \in [0, 5]$ is the score for ECS sub-criterion $i$, $W_i^{(\text{industry})}$ is the industry-specific weight, $n_{\text{ECS}}$ is the number of ECS sub-criteria, and max_score = 5.

*3) Interpretation:* The resulting ECS provides a normalized measure of a dataset's adherence to ethical and privacy standards within specific industry contexts. Healthcare organizations typically emphasize PII handling, financial institutions prioritize regulatory compliance, and government sectors focus on access controls and security protocols. A higher ECS indicates better alignment with ethical considerations for the specified industry, making the dataset more suitable for deployment in regulated environments or sensitive applications within that sector.

### E. Data Quality Score

The Data Quality Score is a measurement that evaluates the inherent characteristics of a dataset that impact the development of ML and artificial intelligence models. Unlike previous dimensions that concentrate on external relevance factors, the DQS focuses on essential attributes that influence a dataset's effectiveness in creating accurate detection models.

*1) Methodology and Quality Dimensions:* The DQS categorizes datasets into three primary quality dimensions:

· **Statistical Properties:** This dimension assesses the statistical integrity and distributional realism of the dataset concerning its application to network traffic analysis. A key factor is class balance, as a significant disparity between minority and majority classes may induce a classification bias towards the dominant class. Such a bias can impair a model's sensitivity to detecting rare but high-impact events, including advanced persistent threats and insider breaches (Ring et al., 2019; Gharib et al., 2016). The investigation also extends to feature distributions, because non-representative or highly skewed data can distort a model's decision boundaries, degrade its calibration, and increase both false positive and false negative rates during operational use. Furthermore, the presence of outliers and noise is considered. While excessive noise can obfuscate underlying data patterns and mislead learning algorithms, it is crucial to preserve legitimate outliers that represent the rare yet significant events requiring comprehensive detection coverage (Kenyon et al., 2020).

TABLE IV
CRITERIA FOR ETHICAL COMPLIANCE SCORE (ECS)

| Criteria | Sub-Criteria | Description | Evaluation Metrics |
|---|---|---|---|
| **1. Network Data Anonymisation** | | | |
| *Evaluation Scale for Sub-criteria 1.1–1.4 (0–5):* 5 = Complete anonymisation without data utility loss; 4 = Comprehensive anonymisation; 3 = Moderate anonymisation; 2 = Limited anonymisation; 1 = Minimal anonymisation efforts; 0 = No anonymisation. | | | |
| | **1.1 PII Removal Techniques** | Identify and remove personally identifiable information (PII) while preserving analytical utility. | Effectiveness of PII/IP anonymisation; completeness of PII removal. |
| | **1.2 IP Address Obfuscation** | Mask or transform IP addresses to mitigate re-identification risk while retaining forensic usefulness. | Preservation of forensic utility; re-identification risk. |
| | **1.3 Payload Sanitisation** | Redact or transform packet payloads to prevent disclosure of sensitive content. | Degree of content protection; retention of analytical value. |
| | **1.4 Metadata Anonymisation** | Protect quasi-identifiers (timestamps, ports, protocol fields) to limit linkage attacks. | Risk of re-identification through metadata. |
| **2. Forensic Integrity Preservation** | | | |
| *Evaluation Scale for Sub-criteria 2.1–2.4 (0–5):* 5 = Exceptional forensic integrity; 4 = Robust integrity preservation; 3 = Adequate integrity measures; 2 = Limited integrity protection; 1 = Minimal forensic considerations; 0 = No integrity mechanisms. | | | |
| | **2.1 Evidence Authenticity** | Guarantee that data originates from the stated source and remains unaltered. | Use of cryptographic verification (hashes, signatures, integrity checks). |
| | **2.2 Data Tampering Prevention** | Detect and record unauthorised modification after capture. | Audit-trail completeness; tamper-evident storage. |
| | **2.3 Chain of Custody** | Document dataset handling and storage across its lifecycle. | Continuity and traceability of custody records. |
| | **2.4 Verifiable Data Provenance** | Trace origin and transformations applied during anonymisation. | Provenance logs; preservation of original context. |
| **3. Consent and Notification Mechanisms** | | | |
| *Evaluation Scale for Sub-criteria 3.1–3.4 (0–5):* 5 = Comprehensive consent framework; 4 = Extensive notification mechanisms; 3 = Moderate consent provisions; 2 = Limited notification; 1 = Minimal consent considerations; 0 = No consent mechanisms. | | | |
| | **3.1 Organisational Consent** | Formal permissions and agreements for data collection, sharing, and use. | Clarity and completeness of monitoring and consent policies. |
| | **3.2 Network Monitoring Disclosure** | Transparency about monitoring scope, purposes, and involved parties. | Explicit consent documentation; public or internal disclosure mechanisms. |
| | **3.3 Usage Limitation** | Restrict use only to specified, legitimate purposes. | Scope and specificity of permitted uses; enforcement mechanisms. |
| | **3.4 Opt-out Provisions** | Allow affected parties to decline collection or specific processing. | Availability and accessibility of opt-out options. |
| **4. Ethical Research Use** | | | |
| *Evaluation Scale for Sub-criteria 4.1–4.4 (0–5):* 5 = Exemplary ethical research framework; 4 = Comprehensive research guidelines; 3 = Adequate research ethics; 2 = Limited research guidelines; 1 = Minimal research guidelines; 0 = No research-use restrictions. | | | |
| | **4.1 Research Purpose Limitations** | Restrict dataset use to legitimate research aims (e.g., IDS/IPS evaluation, security analysis). | Alignment with ethics approvals and stated objectives. |
| | **4.2 Academic Integrity** | Ensure honest reporting, proper citation, and transparency. | Publication and sharing guidelines; citation practices. |
| | **4.3 Security Research Guidelines** | Adhere to responsible security research practices (e.g., coordinated disclosure). | Presence and specificity of research-ethics policies or codes of conduct. |
| | **4.4 Responsible Disclosure** | Minimise harm and ensure coordinated reporting of sensitive findings. | Formal disclosure procedures; ethical-use safeguards. |

· **ML Readiness:** This dimension assesses a dataset's structural and semantic viability for effective ML

model development. A central criterion is feature independence, as high multicollinearity among variables can artificially inflate importance metrics, compromise model interpretability, and destabilize coefficient estimates. Such consequences are particularly detrimental in security applications where model explainability is paramount for forensic analysis (Hesford et al., 2024). Equally vital is label consistency, because mislabeled instances corrupt the ground truth, leading models to learn spurious correlations, which in turn reduces detection accuracy and potentially elevates the rate of false alarms. Furthermore, data completeness is gauged by quantifying missing values and evaluating the suitability of handling techniques. Inadequate management of missing data can introduce bias into statistical relationships and result in instability within the model's predictive outputs (Verma et al., 2019).

· **Model Development Suitability:** The third dimension evaluates the dataset's capacity to support rigorous, unbiased, and reproducible model assessment. A primary consideration is the suitability of the train/test split, with particular emphasis on preventing temporal data leakage and ensuring that both partitions maintain representative distributions of network behaviors (Jindal and Anwar, 2021). Additionally, generalisation indicators are scrutinized to ascertain if models trained on the data exhibit transferability across various timeframes, domains, or operational contexts. This is a critical property for adapting to evolving threat landscapes (Kenyon et al., 2020). Finally, the cross-validation methodology is assessed, prioritizing stratified, temporal, or domain-aware approaches that are known to reduce bias in reported performance metrics and ensure the reproducibility of experimental findings in IDS/IPS evaluation studies (Ring et al., 2019).

This framework ensures that datasets contain the requisite quality features for the development of robust and dependable ML and DL models, irrespective of their specific application context.

*2) Formula and Calculation:* The DQS is calculated using a straightforward normalization formula:

$$\text{DQS} = \frac{\sum_{i=1}^{n_{\text{DQS}}} \text{Score}(c_i)}{n_{\text{DQS}} \times \max(\text{Score})}, \tag{10}$$

where $\text{Score}(c_i) \in [0, 5]$ is the score for DQS subcriterion $i$, $n_{\text{DQS}}$ is the number of DQS sub-criteria, and max_score = 5.

*3) Interpretation:* This calculation yields a normalized score between 0 and 1, with values closer to 1 indicating higher data quality. By focusing exclusively on intrinsic data quality, DQS complements other evaluation frameworks and contributes to a comprehensive understanding of dataset suitability for advanced IDS's. Unlike the other metrics in our framework, DQS is not industry-specific but provides a universal assessment of a dataset's fundamental quality characteristics that affect model development across all application contexts.

## IV. IMPLEMENTATION

To ensure comprehensive validation, we selected ten publicly available IDS/IPS datasets that collectively represent the principal developmental stages and domains of dataset evolution. Network Security Laboratory - Knowledge Discovery and Data Mining (NSL-KDD, 2009) reflects the earliest benchmark attempts to correct flaws in KDD-CUP'99 and remains widely cited as a legacy reference. The University of New South Wales Network Behavior 2015 (UNSW-NB15), Canadian Institute for Cybersecurity Intrusion Detection System 2017 (CICIDS2017) and Canadian Institute for Cybersecurity Intrusion Detection System 2018 (CICIDS2018) represent general-purpose datasets that introduced more realistic traffic and diverse attack types. More recent collections capture domain-specific requirements: Telemetry of Network Internet of Things (ToN-IoT, 2020) addresses distributed IoT environments, Canadian Institute for Cybersecurity Internet of Medical Things (CIC-IoMT, 2024) focuses on medical devices, Edge Internet of Things (Edge-IoT, 2021) and Edge Computing Unit Internet of Healthcare Things (ECU-IoHT, 2020) target edge and healthcare IoT, respectively. Finally, Canadian Institute for Cybersecurity Internet of Vehicles 2024 (CICIoV-24) and Canadian Institute for Cybersecurity UNSW Network Behavior 2015 (CIC-UNSW-NB15, 2024) exemplify the newest generation of hybrid datasets, combining realistic traffic capture with broader attack coverage. Taken together, these datasets span the main categories and developmental trends of current IDS/IPS datasets, ensuring that our evaluation framework is tested across both historical benchmarks and modern, sector-specific collections.

For the implementation, we evaluate each dataset against the five metrics outlined in our framework, with particular focus on the healthcare industry context for the ARS evaluation.

### A. Implementation Methodology

*1) Industry Context Definition:* Our framework utilizes the MITRE ATT&CK Enterprise knowledge base, accessed through their official CTI (Cyber Threat Intelligence) repository(MITRE). We programmatically retrieve the enterprise attack data from the MITRE GitHub repository , ensuring access to the most current threat intelligence. This structured JSON data provides comprehensive information about threat actors, techniques, and their relationships, forming the foundation for our analysis. For the healthcare industry analysis, we identified specific domain-relevant keywords, For example for Healthcare sector: 'healthcare', 'hospital', 'medical', 'pharma', and 'biotech'. For finance sector: 'finance', 'banking', 'bank', 'credit', 'insurance', 'financial'. For the retail sector: 'POS', 'retail', 'Payment Card', 'Customer', 'Supply Chain'. These keywords serve as the foundation for identifying industry-relevant threat actors and their associated techniques within the MITRE ATT&CK framework.

*2) Dataset Mapping Implementation:* After identifying industry-relevant techniques, the next critical step involves mapping dataset attack patterns to MITRE

TABLE V
CRITERIA FOR DATA QUALITY SCORE

| Criteria | Sub-Criteria | Description | Evaluation Metrics | Evaluation Scale (0–5) |
|---|---|---|---|---|
| Statistical Properties | 1.1 Class Balance | Measures the balance between attack and normal traffic classes to avoid bias and ensure rare but important attacks are represented. | Distribution ratio of classes. | **5** – Balanced (0.8–1.0); **4** – Well balanced (0.6–0.79); **3** – Moderate imbalance (0.4–0.59); **2** – Imbalanced (0.2–0.39); **1** – Severe imbalance (0.05–0.19); **0** – Extreme imbalance (< 0.05). |
| | 1.2 Feature Distribution | Evaluates whether features follow expected distributions and represent realistic network behavior. | Normality, skewness, representativeness. | **5** – Features are normally distributed with no significant skewness; **4** – Minor skewness; **3** – Moderate skewness; **2** – Severe skewness in some features; **1** – Severe skewness in most features; **0** – Extreme skewness with non-representative features. |
| | 1.3 Presence of Outliers/Noise | Identifies noise or anomalous points that may distort training, while distinguishing legitimate rare events. | Impact of noise/outliers on model learning. | **5** – Negligible noise (<1%); **4** – Low noise (1–5%); **3** – Moderate noise (6–10%); **2** – High noise (11–15%); **1** – Very high noise (16–20%); **0** – Extreme noise (>20%). |
| ML Readiness | 2.1 Feature Independence | Checks multicollinearity and correlations between features that may undermine model stability. | Pairwise correlation ($|r|$). | **5** – No significant correlation ($|r| < 0.3$); **4** – Low correlation ($0.3 \leq |r| < 0.5$); **3** – Moderate correlation ($0.5 \leq |r| < 0.7$); **2** – High correlation ($0.7 \leq |r| < 0.85$); **1** – Very high correlation ($0.85 \leq |r| < 0.95$); **0** – Extreme correlation ($|r| \geq 0.95$). |
| | 2.2 Label Consistency | Ensures attack/normal labels are accurate and trustworthy as ground truth. | Labelling error rate. | **5** – 0% mislabelled samples; **4** – ≤2% mislabelled samples; **3** – 3–5% mislabelled samples; **2** – 6–10% mislabelled samples; **1** – 11–15% mislabelled samples; **0** – >15% mislabelled samples. |
| | 2.3 Data Completeness | Assesses missing values and whether handling methods prevent bias. | Missingness rate and handling. | **5** – No missing values; **4** – ≤2% missing values handled properly; **3** – 3–5%; **2** – 6–10%; **1** – 11–15%; **0** – >15% missing values with poor handling. |
| Model Development Suitability | 3.1 Train/Test Split Suitability | Evaluates partitioning for temporal validity and class representativeness, avoiding leakage. | Split quality, leakage prevention. | **5** – Well-designed split, no leakage; **4** – Good split with minimal issues; **3** – Acceptable split; **2** – Poor split with some leakage; **1** – Very poor split; **0** – Split with clear leakage. |
| | 3.2 Generalisation Indicators | Checks whether models trained on the dataset generalise across contexts or time periods. | Evidence of transferability. | **5** – Strong evidence of generalisation; **4** – Good evidence; **3** – Moderate evidence; **2** – Limited evidence; **1** – Minimal evidence; **0** – No evidence. |
| | 3.3 Cross-validation Approach | Assesses whether the CV method (e.g., stratified, temporal) is appropriate to reduce bias. | Applicability of CV methodology. | **5** – Appropriate CV with stratification/temporal design; **4** – Good CV; **3** – Basic CV method; **2** – Weak CV setup; **1** – Minimal CV; **0** – No CV. |

ATT&CK techniques. Our framework employs natural language processing techniques to automate this mapping process, ensuring consistency and reproducibility while reducing reliance on manual classification.

The GenerateAttackContext function constructs a comprehensive textual representation of each attack type by analyzing its network behavior characteristics. We extract key statistical features that distinguish each attack type and include protocol and service information to provide context for the semantic matching process. For each relevant feature, we calculate the relative difference from normal traffic patterns, classifying them as "high," "moderate," or "low" based on standardized thresholds.

By encoding these contextual descriptions using a pretrained sentence transformer model, we generate vector representations that capture the semantic essence of each attack type.

*3) Industry-Specific Weighting Derivation for TERS:*
To account for varying technical requirements across industries, we developed an industry-specific weighting system through a three-phase process:

· **Literature Analysis:** We conducted a systematic review of industry-specific cybersecurity publications, including sector-specific security frameworks (National Institute of Standards and Technology, 2018), (Health Information Trust Alliance, 2020),

---

**Algorithm 1:** Industry-Specific Threat Analysis Pseudocode

---

**Input:** MITRE ATT&CK dataset $D$, Industry keywords $K$
**Output:** Normalized weights $W$, Industry-specific technique details $T$
**Initialize:**
- $R \leftarrow \varnothing$ (set of relevant threat groups)
- $C \leftarrow \varnothing$ (counter for technique occurrences)
- $A \leftarrow$ set of active (non-revoked, non-deprecated, and classified as main techniques) techniques

**Identify relevant threat groups: foreach** *threat group* $g \in D$ **do**
    **if** *any $k \in K$ matches $g$'s name or description or aliases or* **then**
        | Add $g$ to $R$
    **end**
**end**
**Calculate technique weights: foreach** *relationship $r \in D$* **do**
    **if** *$r.source \in R$ and $r.target \in A$* **then**
        | Increment $C[r.target]$
    **end**
**end**
$total\_uses \leftarrow \sum C.values()$ $min\_weight \leftarrow 1/|A|$
**foreach** *technique $t \in A$* **do**
    $raw\_weight[t] \leftarrow$ $max(C[t]/total\_uses, min\_weight)$
**end**
**Normalize weights:**
$max\_weight \leftarrow max(raw\_weight.values())$ **foreach** *technique $t \in A$* **do**
    | $W[t] \leftarrow raw\_weight[t]/max\_weight$
**end**
**Generate technique details: foreach** *technique $t \in A$ where $C[t] > 0$* **do**
    $T[t] \leftarrow \{$**name** : $t.name$, **description** : $t.description$, **usage_count** : $C[t]$, **weight** : $W[t]\}$
**end**
**Return:** $W, T$

---

**Algorithm 2:** Attack-to-Technique Mapping Pseudocode

---

**Input:** Dataset $D$, Label column $L$, Normal traffic label $N$, Similarity threshold $\theta_{sim}$, Industry techniques $T$
**Output:** Technique mappings $M$ for each attack type
**Initialize:**
- Load pre-trained sentence transformer model
- $A \leftarrow$ Set of unique attack labels in $D$ where label $\neq N$
- $P \leftarrow$ Protocol information for each attack type
- $S \leftarrow$ Service information for each attack type
- $F \leftarrow$ Relevant features for each attack type
- $M \leftarrow \varnothing$ (mapping results)

**Generate attack context and embeddings: foreach** *attack type $a \in A$* **do**
    $context_a \leftarrow$ GenerateAttackContext($D$, $L$, $a$, $P[a]$, $S[a]$, $F[a]$)
    $E_a \leftarrow$ Encode($context_a$)
**end**
**Generate technique embeddings: foreach** *technique $t \in T$* **do**
    $text_t \leftarrow$ Concatenate($t.id$, $t.name$, $t.description$)
    $E_t \leftarrow$ Encode($text_t$)
**end**
**Perform similarity mapping: foreach** *attack type $a \in A$* **do**
    **foreach** *technique $t \in T$* **do**
        $sim \leftarrow$ CosineSimilarity($E_a$, $E_t$)
        **if** $sim \geq \theta_{sim}$ **then**
            **if** *multiple techniques achieve $sim = \theta_{sim}$* **then**
            | select technique with higher $W_t$;
            **end**
            $M[a] \leftarrow M[a] \cup \{t\}$;
        **end**
    **end**
**end**
**Return:** $M$

---

regulatory guidance documents (Federal Financial Institutions Examination Council, 2015), and empirical studies of attack patterns across different sectors (Hutchins et al., 2010), (National Institute of Standards and Technology, 2023).

· **Expert Consultation:** We consulted with cybersecurity professionals specializing in each target industry (healthcare, finance, government, energy, and retail) through a structured interview process to validate and refine the weights derived from literature.

· **Cross-Validation:** We employed a cross-validation approach where weights were applied to known benchmark datasets to verify that the resulting scores aligned with expert consensus on dataset suitability for each industry.

This process resulted in a comprehensive weighting matrix that reflects the relative importance of each technical characteristic across different industry environments, enabling our framework to provide contextually relevant evaluations.

Table VI presents the industry-specific weights derived from our analysis, reflecting the relative importance of each criterion across different sectors. These weights were normalized to ensure comparability while preserving the relative emphasis within each industry context.

The healthcare sector demonstrates particularly high weights for payload detail (5) and industry-specific protocols (5), reflecting the importance of deep packet inspection for identifying threats to medical systems and the prevalence of specialized medical device protocols (Koutras et al., 2020). In contrast, financial services place greater emphasis on protocol diversity (5) and metadata variety (5), consistent with the sector's complex transaction systems and regulatory requirements for comprehensive monitoring (Financial Services Information Sharing and Analysis Center (FS-ISAC), 2023).

*4) Industry-Specific Weighting Derivation for ECS:*
To account for varying ethical priorities across industries, we developed an industry-specific weighting system through a three-phase process:

· **Regulatory Analysis:** We conducted a systematic review of industry-specific privacy regulations and ethical guidelines, including HIPAA for healthcare (U.S. Department of Health & Human Services, 2013), GLBA for finance (Federal Trade Commission, 2005), and sector-specific provisions of GDPR (European Data Protection Board, 2019) and other

TABLE VI
EVIDENCE-BASED INDUSTRY WEIGHTING FOR TERS

| Criteria | Sub-Criteria | Healthcare | Finance | Retail | Government | Energy | Rationale |
|---|---|---|---|---|---|---|---|
| Network & Protocol Comprehensiveness | Protocol Diversity | 4 | 5 | 3 | 5 | 4 | Financial and government networks require detection across heterogeneous internal/external protocols (Federal Financial Institutions Examination Council, 2015; Financial Services Information Sharing and Analysis Center (FS-ISAC), 2023); retail typically operates with fewer payment-oriented protocols (Financial Services Information Sharing and Analysis Center (FS-ISAC), 2023); healthcare and energy require multi-protocol coverage but are more domain-specific (National Institute of Standards and Technology, 2018; Koutras et al., 2020). |
| | OSI Layer Representation | 3 | 4 | 3 | 5 | 4 | Government and finance require comprehensive coverage for multi-layered enterprise services (Federal Financial Institutions Examination Council, 2015; Financial Services Information Sharing and Analysis Center (FS-ISAC), 2023); healthcare and energy focus on critical layers used in sector systems (National Institute of Standards and Technology, 2018; Koutras et al., 2020). |
| | Inter-protocol Complexity | 3 | 4 | 3 | 4 | 4 | Finance, government, and energy networks handle diverse cross-protocol interactions (National Institute of Standards and Technology, 2018; Federal Financial Institutions Examination Council, 2015; Financial Services Information Sharing and Analysis Center (FS-ISAC), 2023); healthcare and retail have more structured communication patterns (Koutras et al., 2020). |
| Data Capture Methodology | Diversity of Capture Methods | 3 | 3 | 2 | 4 | 4 | Government and energy networks incorporate mixed capture sources (enterprise, OT, cloud) (National Institute of Standards and Technology, 2018, 2006); retail uses more standardised sources (Financial Services Information Sharing and Analysis Center (FS-ISAC), 2023). |
| | Granularity & Temporal Coverage | 4 | 5 | 3 | 5 | 5 | Finance, government, and energy require long-term, varied traffic observation for compliance and resilience (National Institute of Standards and Technology, 2018; Federal Financial Institutions Examination Council, 2015; Financial Services Information Sharing and Analysis Center (FS-ISAC), 2023). |
| | Capture Environment Authenticity | 5 | 4 | 3 | 5 | 5 | Healthcare, government, and energy need live operational traffic for realistic testing (National Institute of Standards and Technology, 2018; Koutras et al., 2020; National Institute of Standards and Technology, 2006). |
| Payload & Metadata Analysis | Payload Detail | 5 | 4 | 3 | 5 | 4 | Healthcare and government require deep packet inspection for regulatory or security compliance (National Institute of Standards and Technology, 2018; Koutras et al., 2020). |
| | Metadata Variety | 4 | 5 | 3 | 4 | 4 | Finance requires extensive metadata for transaction tracing (Federal Financial Institutions Examination Council, 2015; Financial Services Information Sharing and Analysis Center (FS-ISAC), 2023); healthcare and energy require broad network metadata (National Institute of Standards and Technology, 2018; Koutras et al., 2020). |
| | Feature Extraction Potential | 4 | 5 | 3 | 4 | 4 | Finance requires large-scale feature variety for fraud/attack detection (Federal Financial Institutions Examination Council, 2015; Financial Services Information Sharing and Analysis Center (FS-ISAC), 2023); healthcare and energy require detailed technical features (National Institute of Standards and Technology, 2018; Koutras et al., 2020). |
| Recording Environment Characterization | Environment Type | 5 | 3 | 4 | 4 | 5 | Healthcare and energy operate across multiple environment types including OT/IoT and enterprise (National Institute of Standards and Technology, 2018; Koutras et al., 2020); government operates mixed secure networks (National Institute of Standards and Technology, 2006). |
| | Traffic Authenticity | 5 | 5 | 4 | 5 | 5 | Healthcare, finance, government, and energy require realistic patterns for operational model training (National Institute of Standards and Technology, 2018; Federal Financial Institutions Examination Council, 2015; Koutras et al., 2020; National Institute of Standards and Technology, 2006). |
| Specialized Domain Relevance | Industry-Specific Protocols | 5 | 4 | 3 | 4 | 5 | Healthcare uses DICOM, HL7 (Koutras et al., 2020); energy uses Modbus, DNP3 (National Institute of Standards and Technology, 2018); government supports classified/specialised protocols (National Institute of Standards and Technology, 2006). |
| | Specialised Network Configurations | 4 | 4 | 3 | 5 | 5 | Energy, healthcare, and government require complex architectures including OT and cloud-edge (National Institute of Standards and Technology, 2018; Koutras et al., 2020; National Institute of Standards and Technology, 2006). |
| | Security Controls Representation | 5 | 5 | 4 | 5 | 5 | Finance, healthcare, and energy require embedded controls such as encryption and firewalls for compliance (National Institute of Standards and Technology, 2018; Federal Financial Institutions Examination Council, 2015; Koutras et al., 2020). |

privacy frameworks.

- **Expert Consultation:** We consulted with privacy and ethics specialists from each target industry through a structured interview process to validate and refine the weights derived from regulatory analysis.
- **Cross-Validation:** We employed a cross-validation approach where weights were applied to known benchmark datasets to verify that the resulting scores aligned with expert consensus on the ethical considerations for each industry.

This process resulted in a comprehensive weighting matrix that reflects the relative importance of each ethical characteristic across different industry environments, enabling contextually relevant evaluations. The ECS calculation integrates the sub-criteria scores with industry-specific weights in a normalized formula. This section details the calculation methodology and presents the evaluation results. Table VII presents the industry-specific weights derived from our analysis, reflecting the relative importance of ethical considerations across different sectors. These weights were normalized to ensure comparability while preserving the relative emphasis within each industry context.

The healthcare sector demonstrates particularly high weights for PII removal techniques (5), payload sanitization (5), and consent-related criteria (5), reflecting the stringent privacy requirements for protected health information under regulations like HIPAA (U.S. Department of Health & Human Services, 2013). In contrast, government sector places greater emphasis on forensic integrity aspects such as evidence authenticity (5) and chain of custody (5), consistent with legal requirements for admissibility of digital evidence (National Institute of Standards and Technology, 2006).

### B. Score Calculation Implementation

After mapping dataset attacks to MITRE techniques, our framework calculates the five relevance scores. Here we present the pseudocode implementations for ARS and TRS.

*1) ARS:* The ARS provides a weighted measure of how well a dataset covers industry-specific threats, prioritizing techniques that are more frequently used by relevant threat actors. This weighting ensures that datasets scoring highly on ARS will be more effective at detecting the most common attack patterns in the target industry.

*2) TRS:* The TRS employs a logistic-like decay function rather than exponential decay, providing a more gradual deterioration in relevance over time. This approach recognizes that while older datasets and techniques become less relevant, they don't immediately lose all value. The dual temporal consideration of evaluating both dataset age and the currency of included attack techniques, provides a nuanced measure of temporal relevance that reflects real-world security operations.

*3) TERS:* Each of the fifteen sub-criteria in Table III is evaluated strictly in accordance with its operational definition, where each value on the 0–5 scale corresponds to measurable thresholds defined within the framework. For every dataset–industry combination, sub-criterion scores are assigned solely on the basis of verifiable evidence obtained from the dataset's primary publication, official documentation, or independent peer-reviewed analyses. All evidence sources are recorded in an audit table with corresponding reference numbers to enable complete traceability and replication of results. The industry-specific weights presented in Table VI were determined through examination of sector-specific cybersecurity standards, regulatory guidance, and documented attack patterns, supported by established knowledge of operational requirements in each sector. The final TeRS for each dataset–industry pair is calculated

---

**Algorithm 3: ARSCalculation**

**Input:** Dataset techniques $D_T$, Industry techniques $I_T$
**Output:** Attack Relevance Score $ARS$, Coverage statistics
**Initialize:**
- Sort $I_T$ by weight (descending)
- *covered _techniques* ← ∅
- *missing _techniques* ← ∅
- *dataset _ids* ← Set of technique IDs in $D_T$
- *total_weight* ← 0
- *covered _weight* ← 0

**Calculate weighted coverage: foreach** *technique*
$(id, info) \in I_T$ **do**
  *weight* ← *info.weight*
  *total_weight* ← *total_weight + weight* **if**
  *id* ∈ *dataset _ids* **then**
    *covered _weight* ← *covered _weight + weight*
    Add {*id, name, weight*} to *covered _techniques*
  **end**
  **else**
    Add {*id, name, weight*} to *missing _techniques*
  **end**
**end**
**Calculate coverage score: if** *total _weight* > 0 **then**
  $ARS$ ← *covered_weight/total _weight*
**end**
**else**
  $ARS$ ← 0
**end**
**Identify high-priority missing techniques:** *threshold* ←
weight at 75th percentile of $I_T$ *missing _high _priority* ←
Techniques in *missing_techniques* with
*weight ≥ threshold*
**Return:** $ARS$, coverage statistics

---

using Equation 8, in which the scores are combined with the predetermined industry weights in a normalized computation. This procedure ensures that identical input evidence and scoring rules will produce identical results, thereby maintaining objectivity. Each score is supported by documented evidence from the datasets' published descriptions and independent analyses.For example, NSL-KDD's protocol diversity score of 1 is justified by its documented limitation to basic TCP/IP protocols as noted by Tavallaee et al. (Tavallaee et al., 2009), while CIC-IoMT's score of 3 reflects its documented support for Wi-Fi, MQTT, and Bluetooth protocols specifically designed for medical devices as detailed by Dadkhah et al. (Dadkhah et al., 2024). Table X presents the calculated TeRS scores for each dataset across the five industry sectors.

*4) ECS:* Each dataset was evaluated against the 16 sub-criteria using the explicit scoring guidelines, with reference to specific characteristics documented in the literature. Table XI presents these evaluations for nine prominent cybersecurity datasets. Scores are determined exclusively from verifiable information contained in the dataset's official documentation, creator publications, or independent evaluations. Evidence for each score is documented in an audit table with the relevant reference citation, ensuring transparency and enabling independent

TABLE VII
EVIDENCE-BASED INDUSTRY WEIGHTING FOR ECS

| Criteria | Sub-Criteria | Healthcare | Finance | Retail | Government | Energy | Rationale |
|---|---|---|---|---|---|---|---|
| Network Data Anonymization | PII Removal Techniques | 5 | 5 | 4 | 5 | 3 | Healthcare and finance have strict privacy regulations (HIPAA, GLBA) (U.S. Department of Health & Human Services, 2013; Federal Trade Commission, 2005); government sensitive data also requires full anonymisation (National Institute of Standards and Technology, 2006). |
| | IP Address Obfuscation | 4 | 4 | 3 | 5 | 3 | Government systems require complete address protection for operational security (National Institute of Standards and Technology, 2006); healthcare and finance protect client data (U.S. Department of Health & Human Services, 2013; Federal Trade Commission, 2005). |
| Payload Sanitization | Payload Sanitization | 5 | 4 | 3 | 5 | 3 | Healthcare and government require full payload sanitisation for regulatory and security compliance (U.S. Department of Health & Human Services, 2013; National Institute of Standards and Technology, 2006). |
| | Metadata Anonymization | 4 | 5 | 3 | 5 | 3 | Finance and government require protection of metadata to avoid operational compromise (Financial Services Information Sharing and Analysis Center (FS-ISAC), 2023; National Institute of Standards and Technology, 2006). |
| Forensic Integrity Preservation | Evidence Authenticity | 4 | 4 | 3 | 5 | 4 | Government requires admissible digital evidence (National Institute of Standards and Technology, 2006); healthcare and finance require trusted logs for audits (U.S. Department of Health & Human Services, 2013; Federal Trade Commission, 2005). |
| | Data Tampering Prevention | 5 | 3 | 5 | 3 | 4 | Retail requires prevention of transactional record tampering (Financial Services Information Sharing and Analysis Center (FS-ISAC), 2023); healthcare, finance, and energy require similar safeguards (National Institute of Standards and Technology, 2018; U.S. Department of Health & Human Services, 2013; Federal Trade Commission, 2005). |
| | Chain of Custody | 3 | 4 | 2 | 5 | 3 | Government mandates chain of custody for legal admissibility (National Institute of Standards and Technology, 2006); finance and healthcare maintain it for audits (U.S. Department of Health & Human Services, 2013; Federal Trade Commission, 2005). |
| | Verifiable Data Provenance | 3 | 4 | 2 | 5 | 3 | Government requires provenance tracking for security incidents (National Institute of Standards and Technology, 2006); finance and healthcare require it for compliance (U.S. Department of Health & Human Services, 2013; Federal Trade Commission, 2005). |
| Consent and Notification Mechanisms | Organisational Consent | 5 | 4 | 3 | 4 | 3 | Healthcare, finance, and government maintain formal consent frameworks (U.S. Department of Health & Human Services, 2013; Federal Trade Commission, 2005; National Institute of Standards and Technology, 2006). |
| | Network Monitoring Disclosure | 5 | 4 | 3 | 4 | 3 | Transparency in monitoring is regulated in healthcare and finance (U.S. Department of Health & Human Services, 2013; Federal Trade Commission, 2005). |
| | Usage Limitation | 5 | 3 | 4 | 3 | 3 | HIPAA and GLBA impose strict use-limitation clauses (U.S. Department of Health & Human Services, 2013; Federal Trade Commission, 2005). |
| | Opt-out Provisions | 5 | 4 | 3 | 2 | 2 | Healthcare and finance provide opt-out mechanisms under regulatory frameworks (U.S. Department of Health & Human Services, 2013; Federal Trade Commission, 2005). |
| Ethical Research Use | Research Purpose Limitations | 4 | 4 | 3 | 4 | 3 | Academic and applied research restrictions apply in all regulated sectors (U.S. Department of Health & Human Services, 2013; Federal Trade Commission, 2005; National Institute of Standards and Technology, 2006). |
| | Academic Integrity | 4 | 4 | 3 | 4 | 3 | Compliance with integrity standards required for published research in all sectors (U.S. Department of Health & Human Services, 2013; Federal Trade Commission, 2005). |
| | Security Research Guidelines | 4 | 3 | 4 | 5 | 4 | Government emphasises strict vulnerability disclosure processes (National Institute of Standards and Technology, 2006); other sectors follow industry security norms (U.S. Department of Health & Human Services, 2013; Federal Trade Commission, 2005). |
| | Responsible Disclosure | 4 | 3 | 4 | 5 | 4 | Government mandates formal responsible disclosure processes (National Institute of Standards and Technology, 2006); finance and healthcare follow regulated protocols (U.S. Department of Health & Human Services, 2013; Federal Trade Commission, 2005). |

TABLE VIII
ARS ACROSS INDUSTRIES

| Dataset | Year | Healthcare | Finance | Government | Energy | Retail |
|---|---|---|---|---|---|---|
| CIC-UNSW-NB15 | 2024 | 0.4767 | 0.3953 | 0.4626 | 0.5269 | 0.5535 |
| UNSW-15 | 2015 | 0.4025 | 0.3808 | 0.4025 | 0.5027 | 0.5308 |
| CICIoV-24 | 2024 | 0.1145 | 0.0929 | 0.1117 | 0.1285 | 0.3563 |
| ToN-IoT | 2020 | 0.1992 | 0.1614 | 0.1676 | 0.2039 | 0.2259 |
| NSL_KDD | 2009 | 0.1725 | 0.1427 | 0.1426 | 0.1966 | 0.1757 |
| CIC-IoMT | 2024 | 0.4101 | 0.3016 | 0.3140 | 0.3687 | 0.3874 |
| CICIDS2018 | 2018 | 0.3585 | 0.2490 | 0.2696 | 0.3356 | 0.4516 |
| CICIDS2017 | 2017 | 0.1012 | 0.1430 | 0.1070 | 0.1292 | 0.1452 |
| Edge-IOT | 2021 | 0.3882 | 0.2808 | 0.3210 | 0.3694 | 0.4473 |
| ECU-IOHT | 2020 | 0.4642 | 0.3889 | 0.4084 | 0.5046 | 0.5535 |

TABLE IX
TRS ACROSS INDUSTRIES

| Dataset | Year | Healthcare | Finance | Government | Energy | Retail |
|---|---|---|---|---|---|---|
| CIC-UNSW-NB15 | 2024 | 0.0801 | 0.1127 | 0.1165 | 0.0985 | 0.0773 |
| UNSW-15 | 2015 | 0.0618 | 0.0632 | 0.0618 | 0.0665 | 0.2410 |
| CICIoV-24 | 2024 | 0.1393 | 0.1344 | 0.1282 | 0.1291 | 0.2410 |
| ToN-IoT | 2020 | 0.0982 | 0.0952 | 0.1112 | 0.1069 | 0.0974 |
| NSL-KDD | 2009 | 0.0361 | 0.0368 | 0.0416 | 0.0441 | 0.0407 |
| CIC-IMOT | 2024 | 0.0524 | 0.0514 | 0.0523 | 0.0521 | 0.0521 |
| CICIDS2018 | 2018 | 0.0371 | 0.0336 | 0.0392 | 0.0389 | 0.0357 |
| CICIDS2017 | 2017 | 0.0379 | 0.0402 | 0.0470 | 0.0462 | 0.0403 |
| Edge-IOT | 2021 | 0.0899 | 0.0888 | 0.0956 | 0.0932 | 0.0950 |
| ECU-IOHT | 2020 | 0.0920 | 0.1089 | 0.1088 | 0.1046 | 0.0984 |

verification. The industry-specific weights in Table VII reflect the relative importance of different ethical factors across sectors and were determined from established regulatory requirements and operational priorities. These

---

**Algorithm 4:** TRS Calculation

**Input:** attack_technique_map, industry_weights, dataset_year, decay_constant (default: 2.0)

**Output:** trs

**Initialize:**

$current\_year \leftarrow$ current year

$weighted\_relevance\_sum \leftarrow 0$

$weight\_sum \leftarrow 0$

**Calculate dataset age:**

$\Delta t_{\text{dataset}} \leftarrow current\_year - dataset\_year$

**Define temporal decay function:**

$\psi(\Delta t) = \frac{1}{1 + \frac{\Delta t}{decay\_constant}}$

**Calculate dataset decay coefficient:**

$\phi_d \leftarrow \psi(\Delta t_{\text{dataset}})$

**Extract technique IDs from attack_technique_map:**

$technique\_ids \leftarrow$ unique technique IDs from attack_technique_map

**foreach** $technique\ t \in technique\_ids$ **do**

    **Get industry-specific weight:**

    $weight \leftarrow industry\_weights.get(t, 0.1)$

    **Map technique to CVE using SMET:**

    • Extract the technique description

    • Find CVE with highest semantic similarity to the technique description

    • Store mapped CVE ID, CVSS score, and published date

    **if** *CVE found* **then**

        $\Delta t_{\text{tech}} \leftarrow current\_year -$ CVE published year

        $\rho_i \leftarrow \psi(\Delta t_{\text{tech}})$

    **else**

        $relevance\_coefficient \leftarrow 0.5$

    **end**

    $weighted\_relevance\_sum\ +=$

    $weight \times relevance\_coefficient$

    $weight\_sum\ += weight$

**end**

**Calculate final TRS:**

**if** $weight\_sum > 0$ **then**

    $trs \leftarrow \frac{weighted\_relevance\_sum \times \phi_d}{weight\_sum}$

**end**

**return** *trs*

---

TABLE X
FINAL TERS SCORES

| Dataset | Healthcare | Finance | Retail | Government | Energy |
|---|---|---|---|---|---|
| NSL-KDD | 0.016 | 0.016 | 0.020 | 0.015 | 0.015 |
| CICIDS2017 | 0.045 | 0.045 | 0.052 | 0.041 | 0.042 |
| CICIDS2018 | 0.052 | 0.052 | 0.060 | 0.048 | 0.048 |
| ToN-IoT | 0.057 | 0.054 | 0.062 | 0.051 | 0.053 |
| CIC-UNSW-NB15 | 0.069 | 0.067 | 0.077 | 0.062 | 0.064 |
| CICIoV-24 | 0.059 | 0.054 | 0.061 | 0.053 | 0.058 |
| CIC-IoMT | 0.066 | 0.063 | 0.072 | 0.060 | 0.061 |
| Edge-IoT | 0.064 | 0.062 | 0.071 | 0.058 | 0.060 |
| ECU-IoHT | 0.061 | 0.056 | 0.064 | 0.054 | 0.056 |

pre-established weights are applied directly to the scores using Equation (9) to produce a normalized score between 0 and 1. The use of fixed operational definitions, verifiable evidence, and predetermined weighting ensures that ECS outcomes are entirely reproducible when the same rubric and evidence are applied, thereby preserving objectivity in the scoring process.

Each score is supported by documented evidence from the datasets' published descriptions, associated research papers, and independent ethical assessments. For example, NSL-KDD's relatively low score for PII removal techniques (2) reflects its limited anonymization approach as noted by Tavallaee et al. (Tavallaee et al., 2009), while CIC-IoMT's high score (5) is justified by its comprehensive privacy-preserving methodology including full anonymization of medical device identifiers as detailed by Dadkhah et al (Dadkhah et al., 2024). The evaluation reveals significant variations in ethical compliance across datasets. Notably, older datasets like NSL-KDD score poorly on consent mechanisms (0-1 for several criteria), while newer, healthcare-focused datasets like CIC-IoMT and ECU-IoHT demonstrate stronger performance in privacy protection and consent aspects (4-5 for most criteria). Table XI presents the calculated ECS scores for each dataset across the five industry sectors.

*5) DQS:* ach dataset was evaluated against the 9 sub-criteria using the explicit scoring guidelines, with reference to specific characteristics documented in the literature. Table XIII presents these evaluations for nine prominent cybersecurity datasets.

Each score is supported by documented evidence from the datasets' published descriptions and independent analyses. For example, NSL-KDD's class balance score of 1 reflects its documented significant imbalance between attack types, particularly for R2L and U2R attacks, as noted by Tavallaee et al. (Tavallaee et al., 2009). In contrast, CIC-UNSW-NB15 earns a score of 3 for class balance due to its intentionally designed class distribution with comprehensive documentation of class ratios and justification for the chosen distribution, as detailed by Moustafa and Slay (Kasongo and Sun, 2020). The evaluation reveals notable variations in data quality across datasets. Older datasets like NSL-KDD show limitations in multiple quality aspects, while newer datasets like CIC-UNSW-NB15 demonstrate more rigorous attention to quality factors that impact ML model development.

Table XIV presents the calculated Final DQS scores using equation10 for each dataset with the

## V. RESULTS AND ANALYSIS

The outputs of the proposed evaluation framework extend beyond static assessment and serve as actionable inputs for DL-based intrusion detection workflows. Building on established evaluation frameworks for intrusion detection datasets(Gharib et al., 2016), our approach ensures that selected datasets possess the statistical balance, feature independence, and completeness required for effective machine learning model development. As demonstrated in comprehensive surveys of network-based intrusion detection datasets [2], these intrinsic quality characteristics significantly impact model performance, particularly when combined with appropriate feature selection methodologies (Kasongo and Sun, 2020). The framework's ARS and TRS components address critical challenges in dataset fitness for real-world deployment (Kenyon et al., 2020), providing industry-specific threat coverage and temporal currency indicators essential for constructing representative training datasets. This temporal awareness is particularly crucial given empirical evidence that machine learning-based intrusion detection models suffer from limited temporal validity and declining effectiveness over time (Viegas et al., 2020; dos Santos et al., 2022). By prioritizing datasets with high ARS and TRS values, practitioners can mitigate the risk of training on outdated or non-representative attack patterns, thereby improving model generalization and extending operational lifespans in evolving threat environments. Table XV presents the consolidated scores for each evaluated dataset across all dimensions for five industry contexts. The results reveal significant variations in dataset suitability across different evaluation metrics and sectors. Among the ten evaluated datasets, recent collections generally demonstrate higher scores across most metrics. CIC-UNSW-NB15 (2024) achieved the highest composite scores across all industries, particularly for Retail (0.338) and Energy (0.334) sectors, excelling in Data Quality (0.926) and Attack Relevance. UNSW-NB15, despite being published in 2015, showed exceptionally strong performance for the Retail sector (0.366 combined score), driven by its high ARS (0.531) and TRS (0.241). Specialized datasets demonstrated targeted strengths in specific sectors. ECU-IoHT showed particularly strong performance for Healthcare and Retail industries, with ARS values of 0.464 and 0.554 respectively. CIC-IoMT, designed specifically for medical IoT devices, achieved a healthcare ARS of 0.410, significantly outperforming general-purpose datasets in this context. Edge-IoT similarly demonstrated strong healthcare relevance (ARS: 0.388), reflecting its focus on IoT environments common in modern healthcare settings. In contrast, older and more general datasets showed significant deficiencies. NSL-KDD (2009) performed poorly across all dimensions and industries, with particularly low scores in Technical Environment Relevance (0.015-0.020) and Ethical Compliance (0.015-0.021). demonstrated low ARS's across all sectors (ranging from 0.101 to 0.145), suggesting limited coverage of modern attack techniques despite its more recent publication date. CICIoV-24, despite being a 2024 dataset, showed surprisingly low ARS values for most sectors except retail (0.356), demonstrating that publication date alone does not guarantee alignment with industry-specific threat landscapes. ToN-IoT showed moderate performance, with ARS scores ranging from 0.161 (Finance) to 0.226 (Retail), positioning it in the middle tier among evaluated datasets.

TABLE XI
DATASET EVALUATION WITH ETHICAL CONSIDERATIONS

| Criteria | Sub-Criteria | NSL-KDD | CICIDS2017 | CICIDS2018 | TON-IoT | CIC-UNSW-NB15 | CICIoV-24 | CIC-IoMT | Edge-IoT | ECU-IoHT |
|---|---|---|---|---|---|---|---|---|---|---|
| **Network Data Anonymization** | PII Removal Techniques | 2 | 3 | 3 | 4 | 4 | 5 | 4 | 5 | 4 |
| | IP Address Obfuscation | 3 | 3 | 3 | 4 | 4 | 4 | 4 | 4 | 4 |
| | Payload Sanitization | 1 | 3 | 4 | 4 | 5 | 4 | 5 | 4 | 5 |
| **Metadata Anonymization** | Forensic Integrity Preservation | 2 | 3 | 3 | 4 | 4 | 4 | 4 | 4 | 4 |
| | Evidence Authenticity | 2 | 3 | 3 | 4 | 4 | 4 | 4 | 4 | 4 |
| **Data Tampering Prevention** | Data Tampering Prevention | 1 | 2 | 3 | 3 | 4 | 3 | 4 | 3 | 4 |
| | Chain of Custody | 1 | 2 | 3 | 3 | 4 | 3 | 4 | 3 | 4 |
| | Verifiable Data Provenance | 1 | 2 | 3 | 3 | 4 | 3 | 4 | 3 | 4 |
| **Consent and Notification Mechanisms** | Organizational Consent | 1 | 3 | 4 | 4 | 5 | 4 | 5 | 4 | 5 |
| | Network Monitoring Disclosure | 1 | 3 | 4 | 4 | 5 | 4 | 5 | 4 | 5 |
| | Usage Limitation | 2 | 3 | 4 | 4 | 5 | 4 | 5 | 4 | 5 |
| | Opt-out Provisions | 0 | 2 | 3 | 3 | 4 | 3 | 4 | 3 | 4 |
| **Ethical Research Use** | Research Purpose Limitations | 2 | 3 | 4 | 4 | 4 | 4 | 4 | 4 | 4 |
| | Academic Integrity | 3 | 3 | 4 | 4 | 4 | 4 | 4 | 4 | 4 |
| | Security Research Guidelines | 2 | 3 | 4 | 4 | 4 | 4 | 5 | 4 | 5 |
| | Responsible Disclosure | 2 | 3 | 4 | 4 | 4 | 4 | 5 | 4 | 5 |

TABLE XII
FINAL ECS SCORE

| Dataset | Healthcare | Finance | Retail | Government | Energy |
|---|---|---|---|---|---|
| NSL-KDD | 0.017 | 0.017 | 0.021 | 0.015 | 0.020 |
| CICIDS2017 | 0.033 | 0.033 | 0.038 | 0.030 | 0.037 |
| CICIDS2018 | 0.037 | 0.037 | 0.043 | 0.034 | 0.042 |
| ToN-IoT | 0.045 | 0.044 | 0.051 | 0.041 | 0.050 |
| CIC-UNSW-NB15 | 0.050 | 0.049 | 0.057 | 0.045 | 0.055 |
| CICIoV-24 | 0.051 | 0.050 | 0.058 | 0.047 | 0.057 |
| CIC-IoMT | 0.058 | 0.056 | 0.066 | 0.053 | 0.064 |
| Edge-IoT | 0.050 | 0.050 | 0.058 | 0.046 | 0.056 |
| ECU-IoHT | 0.059 | 0.058 | 0.067 | 0.054 | 0.066 |

TABLE XIII
Dataset Evaluation with Documented Evidence for Data Quality

| Criteria | Sub-Criteria | NSL-KDD | CICIDS2017 | CICIDS2018 | TON-IoT | CIC-UNSW-NB15 | CICIoV-24 | CIC-IoMT | Edge-IoT | ECU-IoHT |
|---|---|---|---|---|---|---|---|---|---|---|
| **Statistical Properties** | Class Balance | 1 | 2 | 2 | 3 | 2 | 2 | 2 | 2 | 1 |
| | Feature Distribution | 1 | 2 | 2 | 3 | 2 | 2 | 2 | 2 | 2 |
| | Presence of Outliers/Noise | 1 | 2 | 2 | 2 | 2 | 2 | 2 | 2 | 1 |
| **ML Readiness** | Feature Independence | 1 | 2 | 2 | 3 | 2 | 2 | 2 | 2 | 2 |
| | Label Consistency | 2 | 2 | 3 | 3 | 3 | 3 | 3 | 3 | 2 |
| **Model Development Suitability** | Train/Test Split Suitability | 2 | 2 | 2 | 3 | 2 | 2 | 2 | 2 | 2 |
| | Generalization Indicators | 1 | 2 | 2 | 3 | 2 | 2 | 2 | 2 | 1 |
| | Cross-validation Approach | 1 | 2 | 2 | 3 | 2 | 2 | 2 | 2 | 1 |

TABLE XIV
DQS for Each Dataset

| Dataset | DQS Score |
|---|---|
| NSL-KDD | 0.370 |
| CICIDS2017 | 0.667 |
| CICIDS2018 | 0.704 |
| ToN-IoT | 0.704 |
| CIC-UNSW-NB15 | 0.926 |
| CICIoV-24 | 0.704 |
| CIC-IoMT | 0.704 |
| Edge-IoT | 0.704 |
| ECU-IoHT | 0.519 |

### A. Industry-Specific Attack Relevance Analysis

The ARS revealed significant variations in dataset suitability across different industry sectors, highlighting the importance of industry-specific dataset selection when evaluating IDS/IPS systems. Figure 4 illustrates the comparative ARS values for each dataset across the five industry sectors evaluated in our framework. Based on our comprehensive analysis of industry-specific threat actors and their associated techniques within the MITRE ATT&CK framework, we identified the most critical techniques for each sector.

For each industry sector, we observed specific patterns in dataset performance:

**Healthcare sector**: CIC-UNSW-NB15 (0.477), ECU-IoHT (0.464), and CIC-IoMT (0.410) achieved the highest ARS values, with Edge-IoT (0.388) following closely. These datasets demonstrated strong coverage of healthcare-relevant techniques, particularly PowerShell execution and ingress tool transfer methods commonly used in healthcare-targeted attacks. CICIDS2017 performed poorly (0.101), indicating limited coverage of techniques relevant to modern healthcare threats.

**Finance sector**: All datasets showed relatively lower ARS values for finance compared to other sectors, with CIC-UNSW-NB15 (0.395) and ECU-IoHT (0.389) performing best. This suggests a significant gap in coverage for financial sector-specific attack techniques across all evaluated datasets. The financial sector's emphasis on malicious file execution (T1204.002) and spearphishing attachments (T1566.001) was best represented in these datasets but even the highest scores indicate only moderate alignment with finance-specific threats.

**Retail sector**: ECU-IoHT and CIC-UNSW-NB15 (both 0.554) achieved the highest ARS values for retail, followed by UNSW-NB15 (0.531). CICIoV-24 showed its strongest performance in this sector (0.356), despite relatively weak scores in other industries. This suggests that retail-specific attack techniques, particularly those involving PowerShell and file deletion, are well-represented across multiple datasets.

**Government sector**: CIC-UNSW-NB15 (0.463) and ECU-IoHT (0.408) provided the best coverage of government-relevant techniques, with UNSW-NB15

TABLE XV
Comprehensive Industry-Specific Evaluation Framework for Intrusion Detection Datasets

| Dataset | Industry | Evaluation Metrics | | | | | Combined Score | Key Observations |
|---------|----------|-----|-----|------|-----|-----|--------|-----|
| | | ARS | TRS | TeRS | ECS | DQS | | |
| NSL-KDD (2009) | Healthcare | 0.173 | 0.036 | 0.016 | 0.017 | 0.370 | 0.122 | Demonstrates significant deficiencies across multiple metrics, particularly in technical environment relevance and ethical compliance. Limited representation of modern threat vectors critical to all sectors. |
| | Finance | 0.143 | 0.037 | 0.016 | 0.017 | 0.370 | 0.117 | |
| | Retail | 0.176 | 0.041 | 0.020 | 0.021 | 0.370 | 0.126 | |
| | Gov & Def | 0.143 | 0.042 | 0.015 | 0.015 | 0.370 | 0.117 | |
| | Energy | 0.197 | 0.044 | 0.015 | 0.020 | 0.370 | 0.129 | |
| CICIDS2017 (2017) | Healthcare | 0.101 | 0.038 | 0.045 | 0.033 | 0.667 | 0.177 | Shows improved data quality over NSL-KDD but still exhibits limited alignment with industry-specific threats, particularly for healthcare (0.101 ARS). Moderate ethical compliance improvements. |
| | Finance | 0.143 | 0.040 | 0.045 | 0.033 | 0.667 | 0.186 | |
| | Retail | 0.145 | 0.040 | 0.052 | 0.038 | 0.667 | 0.188 | |
| | Gov & Def | 0.107 | 0.047 | 0.041 | 0.030 | 0.667 | 0.178 | |
| | Energy | 0.129 | 0.046 | 0.042 | 0.037 | 0.667 | 0.184 | |
| CICIDS2018 (2018) | Healthcare | 0.359 | 0.037 | 0.052 | 0.037 | 0.704 | 0.238 | Substantial improvement in attack relevance, especially for retail (0.452) and healthcare (0.359) sectors. Notable improvement in data quality while maintaining consistent technical relevance. |
| | Finance | 0.249 | 0.034 | 0.052 | 0.037 | 0.704 | 0.215 | |
| | Retail | 0.452 | 0.036 | 0.060 | 0.043 | 0.704 | 0.259 | |
| | Gov & Def | 0.270 | 0.039 | 0.048 | 0.034 | 0.704 | 0.219 | |
| | Energy | 0.336 | 0.039 | 0.048 | 0.042 | 0.704 | 0.234 | |
| UNSW-NB15 (2015) | Healthcare | 0.403 | 0.062 | 0.069 | 0.050 | 0.926 | 0.302 | Exceptionally strong data quality (0.926) and high attack relevance across all sectors. Particularly strong for retail (0.531 ARS) and energy (0.503 ARS) sectors despite its 2015 publication date. |
| | Finance | 0.381 | 0.063 | 0.067 | 0.049 | 0.926 | 0.297 | |
| | Retail | 0.531 | 0.241 | 0.077 | 0.057 | 0.926 | 0.366 | |
| | Gov & Def | 0.403 | 0.062 | 0.062 | 0.045 | 0.926 | 0.300 | |
| | Energy | 0.503 | 0.067 | 0.064 | 0.055 | 0.926 | 0.323 | |
| CIC-UNSW-NB15 (2024) | Healthcare | 0.477 | 0.080 | 0.069 | 0.050 | 0.926 | 0.320 | Highest combined scores across most industries, with exceptional attack relevance for retail (0.554) and energy (0.527) sectors. Balances strong temporal relevance with excellent data quality metrics. |
| | Finance | 0.395 | 0.113 | 0.067 | 0.049 | 0.926 | 0.310 | |
| | Retail | 0.554 | 0.077 | 0.077 | 0.057 | 0.926 | 0.338 | |
| | Gov & Def | 0.463 | 0.117 | 0.062 | 0.045 | 0.926 | 0.323 | |
| | Energy | 0.527 | 0.099 | 0.064 | 0.055 | 0.926 | 0.334 | |
| CICIoV-24 (2024) | Healthcare | 0.115 | 0.139 | 0.059 | 0.051 | 0.704 | 0.214 | Despite recent publication (2024), shows uneven attack relevance across sectors with notably higher performance for retail (0.356 ARS) but lower scores for other industries. Strong temporal relevance (TRS). |
| | Finance | 0.093 | 0.134 | 0.054 | 0.050 | 0.704 | 0.207 | |
| | Retail | 0.356 | 0.241 | 0.061 | 0.058 | 0.704 | 0.284 | |
| | Gov & Def | 0.112 | 0.128 | 0.053 | 0.047 | 0.704 | 0.209 | |
| | Energy | 0.129 | 0.129 | 0.058 | 0.057 | 0.704 | 0.215 | |
| CIC-IoMT (2024) | Healthcare | 0.410 | 0.052 | 0.066 | 0.058 | 0.704 | 0.258 | Strong healthcare-specific performance (0.410 ARS) with the highest ethical compliance score (0.058) for healthcare applications, reflecting its specialized focus on medical IoT environments. |
| | Finance | 0.302 | 0.051 | 0.063 | 0.056 | 0.704 | 0.235 | |
| | Retail | 0.387 | 0.052 | 0.072 | 0.066 | 0.704 | 0.256 | |
| | Gov & Def | 0.314 | 0.052 | 0.060 | 0.053 | 0.704 | 0.237 | |
| | Energy | 0.369 | 0.052 | 0.061 | 0.064 | 0.704 | 0.250 | |
| Edge-IoT (2021) | Healthcare | 0.388 | 0.090 | 0.064 | 0.050 | 0.704 | 0.259 | Good attack relevance across sectors, particularly for retail (0.447) and healthcare (0.388). Balanced performance across all evaluation metrics with strong TRSs. |
| | Finance | 0.281 | 0.089 | 0.062 | 0.050 | 0.704 | 0.237 | |
| | Retail | 0.447 | 0.095 | 0.071 | 0.058 | 0.704 | 0.275 | |
| | Gov & Def | 0.321 | 0.096 | 0.058 | 0.046 | 0.704 | 0.245 | |
| | Energy | 0.369 | 0.093 | 0.060 | 0.056 | 0.704 | 0.256 | |
| ToN-IoT (2020) | Healthcare | 0.199 | 0.098 | 0.057 | 0.045 | 0.704 | 0.221 | Moderate ARS's but strong temporal relevance, particularly for government applications (0.111 TRS). Consistent data quality but limited industry-specific technical relevance. |
| | Finance | 0.161 | 0.095 | 0.054 | 0.044 | 0.704 | 0.212 | |
| | Retail | 0.226 | 0.097 | 0.062 | 0.051 | 0.704 | 0.228 | |
| | Gov & Def | 0.168 | 0.111 | 0.051 | 0.041 | 0.704 | 0.215 | |
| | Energy | 0.204 | 0.107 | 0.053 | 0.050 | 0.704 | 0.224 | |
| ECU-IoHT (2020) | Healthcare | 0.464 | 0.092 | 0.061 | 0.059 | 0.519 | 0.239 | Excellent healthcare (0.464) and retail (0.554) attack relevance with strong ethical compliance scores. Lower data quality (0.519) compared to other recent datasets but compensated by strong industry-specific alignment. |
| | Finance | 0.389 | 0.109 | 0.056 | 0.058 | 0.519 | 0.226 | |
| | Retail | 0.554 | 0.098 | 0.064 | 0.067 | 0.519 | 0.260 | |
| | Gov & Def | 0.408 | 0.109 | 0.054 | 0.054 | 0.519 | 0.229 | |
| | Energy | 0.505 | 0.105 | 0.056 | 0.066 | 0.519 | 0.250 | |

Note: ARS = Attack Relevance Score (alignment with industry-specific threats); TRS = Temporal Relevance Score (dataset currency); TeRS = Technical Environment Relevance Score (compatibility with industry technical requirements); ECS = Ethical Compliance Score (adherence to privacy and ethical standards); DQS = Data Quality Score (intrinsic dataset characteristics for ML/AI model development).

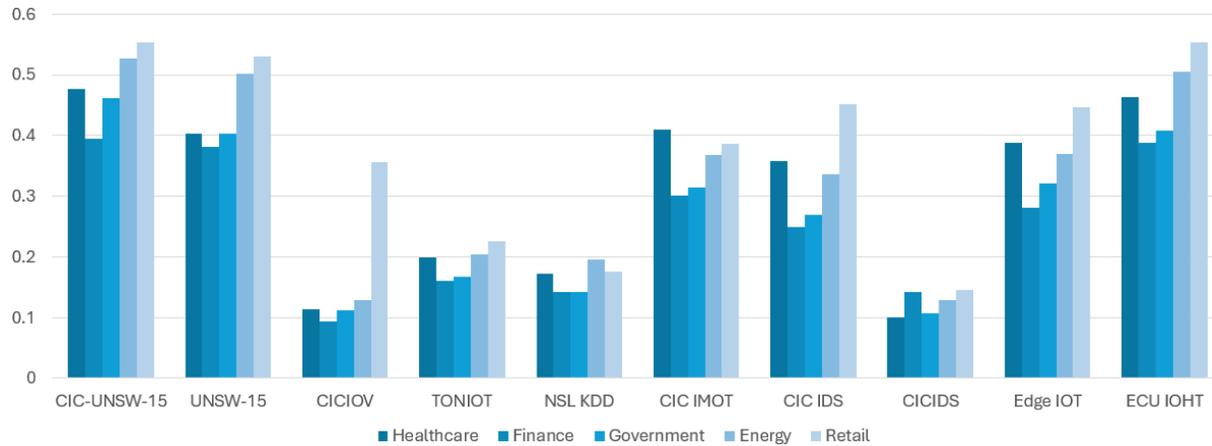

Fig. 4. Comparative ARS values for each dataset across the five industry sectors evaluated in our framework.

(0.403) showing comparable performance. ToN-IoT (0.168) and CICIDS2017 (0.107) demonstrated particularly poor alignment with government threat landscapes, covering less than 20% of relevant techniques.

**Energy sector**: CIC-UNSW-NB15 (0.527) and ECU-IoHT (0.505) significantly outperformed other datasets for the relevance of the energy sector. UNSW-NB15 (0.503) also showed strong performance, suggesting good coverage of techniques like malicious file execution and ingress tool transfer that are prevalent in energy sector attacks. NSL-KDD (0.197) and CICIDS2017 (0.129) showed the weakest alignment with the threats to the energy sector.

### B. Temporal Relevance Analysis

TRS provided crucial insight into the currency of attack patterns from the dataset relative to evolving threat landscapes. Table IX presents these scores across all sectors and datasets. Several notable patterns emerged from this analysis: UNSW-NB15, despite being published in 2015, achieved the highest TRS for the retail sector (0.241), suggesting that its attack techniques have maintained remarkable relevance in this specific context. This finding challenges the assumption that older datasets are universally less temporally relevant and highlights the importance of sector-specific evaluation. CICIoV-24 demonstrated consistently high TRS values across sectors (0.129-0.241), reflecting its recent publication and incorporation of contemporary attack techniques. Similarly, CIC-UNSW-NB15 showed strong temporal relevance for government (0.117) and finance (0.113) sectors. The oldest dataset, NSL-KDD (2009), predictably showed the lowest TRS values across all sectors (0.036-0.044), indicating its attack patterns have significantly diminished in relevance over time. CICIDS2017 and CICIDS2018 also demonstrated low temporal relevance (TRS ranges of 0.034-0.047), despite their more recent publication dates. Edge-IoT and ToN-IoT showed moderate temporal relevance (TRS respective ranges of 0.089-0.096 and 0.095-0.111), positioning them in the middle tier. This suggests these datasets incorporate attack techniques that have maintained reasonable currency despite ongoing threat evolution.

### C. Technical Environment and Ethical Compliance Analysis

The TeRS and ECS scores revealed important insights about dataset applicability to real-world deployment contexts and adherence to privacy and ethical standards. For TeRS, CIC-UNSW-NB15 achieved the highest scores across all sectors (0.062-0.077), with particularly strong performance in retail environments (0.077). This reflects its comprehensive protocol coverage, diverse capture methods, and strong metadata representation. CIC-IoMT demonstrated the second-highest TeRS for healthcare (0.066), highlighting its specialized design for medical device environments. In contrast, NSL-KDD showed exceptionally poor technical environment relevance (0.015-0.020), reflecting its limited protocol diversity and outdated network traffic characteristics. The ECS evaluation revealed considerable variation in ethical compliance across datasets. CIC-IoMT achieved the highest scores across all sectors (0.053-0.066), with particularly strong performance in retail (0.066) and energy (0.064) sectors. ECU-IoHT followed closely (0.054-0.067), demonstrating strong ethical compliance, particularly for real applications. These newer, specialized datasets incorporate more robust privacy protection mechanisms, including comprehensive PII removal and payload sanitization. Older datasets showed significant ethical deficiencies, with NSL-KDD scoring lowest across all sectors (0.015-0.021). CICIDS2017 and CICIDS2018 achieved only moderate ECS values (ranges of 0.030-0.038 and 0.034-0.043, respectively), indicating limited attention to consent mechanisms and data privacy protections. This progression reflects the increasing emphasis on ethical considerations in dataset development over time, particularly for specialized applications like healthcare.

TABLE XVI
COMPREHENSIVE DATASET COMPARISON

| Dataset | Years Since Publication | Best Industry Fit | Highest ARS | Average DQS | Key Strength | Key Limitation |
|---|---|---|---|---|---|---|
| CIC-UNSW-NB15 | 1 (2024) | Retail/Energy | 0.554 (Retail) | 0.926 | Balanced performance across all metrics | Moderate ethical compliance |
| UNSW-NB15 | 10 (2015) | Retail | 0.531 (Retail) | 0.926 | Strong temporal relevance despite age | Limited protocol coverage |
| CICIoV-24 | 1 (2024) | Retail | 0.356 (Retail) | 0.704 | Strong temporal relevance | Low attack relevance for non-retail sectors |
| ToN-IoT | 5 (2020) | Retail | 0.226 (Retail) | 0.704 | Moderate temporal relevance | Limited attack technique coverage |
| NSL-KDD | 16 (2009) | Energy | 0.197 (Energy) | 0.370 | Historical benchmark value | Poor performance across all metrics |
| CIC-IoMT | 1 (2024) | Healthcare | 0.410 (Healthcare) | 0.704 | Strong healthcare protocol coverage | Limited cross-sector applicability |
| CICIDS2018 | 7 (2018) | Retail | 0.452 (Retail) | 0.704 | Good attack relevance for retail | Weak temporal relevance |
| CICIDS2017 | 8 (2017) | Retail | 0.145 (Retail) | 0.667 | Widely used benchmark | Poor attack relevance across all sectors |
| Edge-IoT | 4 (2021) | Retail | 0.447 (Retail) | 0.704 | Strong IoT environment representation | Limited coverage of non-IoT attacks |
| ECU-IoHT | 2 (2020) | Retail | 0.554 (Retail) | 0.519 | Strong ethical compliance | Lower data quality |

### D. Data Quality Analysis

The DQS evaluation focused on intrinsic dataset characteristics that impact ML model development. Table XIV presents these scores, revealing significant variations in fundamental quality attributes: CIC-UNSW-NB15 and UNSW-NB15 achieved the highest DQS (both 0.926), indicating exceptional statistical properties, ML readiness, and model development suitability. These datasets demonstrate strong class balance, feature independence, and cross-validation potential, making them particularly valuable for developing robust ML/AI models regardless of application context. Most mid-range datasets, including CICIDS2017, CICIDS2018, ToN-IoT, CICIoV-24, CIC-IoMT, and Edge-IoT, achieved identical DQS values (0.704). This suggests a standardization of basic data quality practices in contemporary dataset development, with similar attention to feature distribution, noise management, and label consistency. NSL-KDD showed the poorest data quality (0.370), with particular deficiencies in class balance, feature independence, and cross-validation approach. ECU-IoHT demonstrated moderate data quality (0.519), primarily limited by class imbalance issues and the presence of outliers affecting model generalization. These findings highlight that while newer datasets generally demonstrate better data quality characteristics, publication date alone does not guarantee superior quality. The notably high DQS of UNSW-NB15, despite its 2015 publication date, underscores the importance of rigorous dataset design principles regardless of when the dataset was created.

### E. Practical Implications

Our comprehensive evaluation framework yields several important implications for IDS/IPS development and deployment across different industry sectors:

1) **Industry-Specific Dataset Selection**: Organizations should prioritize datasets with high ARS values for their specific sector. Healthcare organizations should consider CIC-UNSW-NB15 (ARS: 0.477) or ECU-IoHT (ARS: 0.464) for general network protection, supplemented with CIC-IoMT (ARS: 0.410) for medical device environments. Financial institutions should select CIC-UNSW-NB15 (ARS: 0.395), though they should be aware of the general gap in finance-specific attack coverage across all datasets. Retail organizations have the widest selection of suitable datasets, with CIC-UNSW-NB15, ECU-IoHT (both ARS: 0.554), and UNSW-NB15 (ARS: 0.531) all demonstrating strong alignment with retail threat landscapes.

2) **Multi-Dataset Training Strategy**: Given that no single dataset excels across all evaluation metrics, organizations should consider using complementary datasets to overcome individual limitations. For example, combining CIC-UNSW-NB15 (strong in attack relevance and data quality) with ECU-IoHT (strong in ethical compliance) would provide more comprehensive coverage for healthcare environments.

3) **Temporal Refresh Considerations**: Systems trained on datasets with low TRS values should be refreshed more frequently. NSL-KDD-based systems (TRS: 0.036-0.044) require frequent retraining, while those using UNSW-NB15 for retail applications (TRS: 0.241) may need less frequent updates. This differential approach to model maintenance can optimize resource allocation while maintaining the detection effectiveness.

4) **Ethical and Privacy Focus**: As regulatory requirements for data privacy increase, particularly in healthcare and finance, organizations should prioritize datasets with higher ECS values. CIC-IoMT (ECS: 0.053-0.066) and ECU-IoHT (ECS: 0.054-0.067) demonstrate stronger ethical compliance than older datasets, making them more suitable for deployment with strict privacy requirements.

5) **Dataset Development Priorities**: Our analysis identifies critical gaps in existing datasets that should guide future development efforts. Notably, there is a significant need for datasets with better coverage of financial sector-specific attack techniques, as even the best-performing dataset achieved an ARS of only 0.395 for this sector.

By applying these insights, organizations can make more informed decisions regarding IDS/IPS dataset selection, training methodologies, and operational deployment, ultimately enhancing their security posture against industry-specific threats.

## VI. BOUNDARIES OF APPLICABILITY AND LIMITATIONS

While the proposed framework offers a structured and comprehensive approach to evaluating the suitability of datasets for IDS/IPS systems, it has application boundaries and limitations. One key constraint lies in the reliance on natural language processing (NLP) techniques, such as cosine similarity, to map attack labels to MITRE ATT&CK techniques. As noted by Steck et al. (Mikolov et al., 2013) and demonstrated in our evaluation of seventeen sentence embedding models (II), cosine similarity may not fully capture semantic nuances, potentially leading to inaccuracies in aligning dataset attacks with their corresponding techniques. Our results showed that even the best-performing model (`acedev003/gte-small-mitre`) achieved only 78.57% overall accuracy, indicating substantial room for improvement in semantic mapping precision. This limitation is particularly evident when attack descriptions contain domain-specific terminology or when techniques share similar surface-level descriptions but differ fundamentally in their implementation or impact.

The framework's industry-specific weights, while derived through expert consultations and cross-validation as outlined in Section IV-A1, may introduce subjectivity as ethical priorities and technical requirements vary significantly across sectors. The weighting matrices presented in VI and VII reflect generalisations that may not fully capture the heterogeneity within industries. Different organizations within the same sector may have varying risk tolerances, regulatory interpretations, and operational priorities, which could affect the applicability of standardised industry weights. This subjectivity may limit the framework's universal applicability within heterogeneous industry segments.

The Temporal Relevance Score (TRS), which employs a fixed decay constant ($\lambda = 2.0$) for threat evolution as described in Section III-B, may oversimplify the dynamic nature of threat landscapes. This limitation is particularly pronounced in industries with rapidly evolving attack patterns, such as healthcare or energy, where the half-life of threat relevance may vary considerably across different attack types and technological contexts. Our empirical analysis (Table IX) revealed that temporal relevance patterns do not follow uniform decay across all sectors, suggesting that the fixed logistic decay function may not adequately capture the non-linear patterns of threat evolution observed in certain industries.

The evaluated datasets themselves often exhibit gaps in coverage, particularly with respect to modern environments such as IoT, cloud scalability, and zero-day threats. As evidenced by our comprehensive evaluation (Table XV), even the highest-scoring datasets like CIC-UNSW-NB15 and UNSW-NB15 demonstrated limited coverage of emerging attack vectors. These limitations reflect broader challenges in dataset design rather than inherent flaws in the framework itself but they constrain the framework's ability to provide comprehensive evaluations for contemporary threat landscapes. The scarcity of high-quality datasets representing emerging attack vectors limits the framework's applicability to cutting-edge security scenarios.

The framework's reliance on existing threat intelligence repositories, particularly the MITRE ATT&CK framework and CVE databases as described in our methodology (Algorithm 1), introduces potential biases toward documented and formally recognised threats.

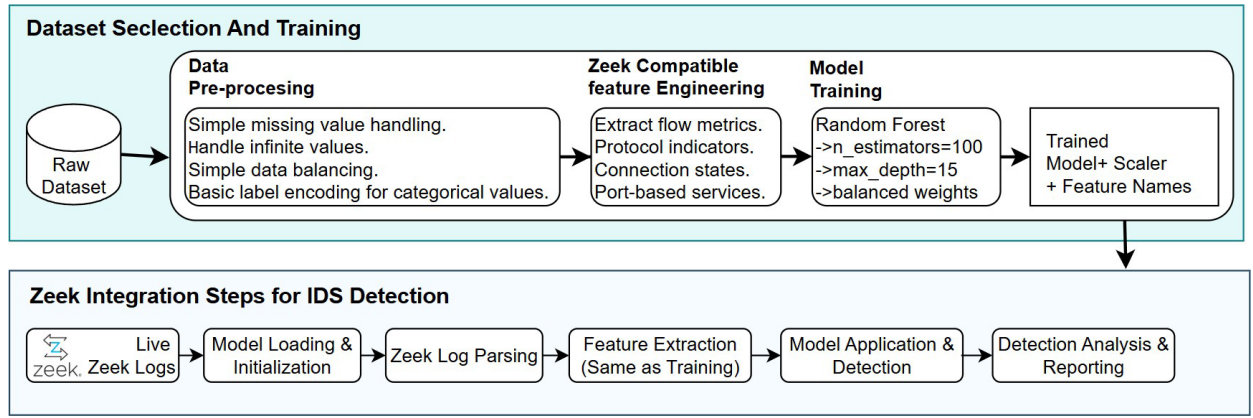

Fig. 5. Overview of the experimental workflow for the empirical validation pipeline.

This dependency may result in under-representation of novel, undisclosed, or sector-specific attack techniques that have not yet been catalogued in these repositories. Our industry-specific threat analysis identified varying numbers of relevant techniques across sectors (ranging from healthcare to retail) but this variation may reflect repository completeness rather than real threat distribution.

## VII. EMPIRICAL VALIDATION WITH CASE STUDY

This case study examines whether the proposed dataset-evaluation framework anticipates operational behavior when two classifiers, trained on different datasets under a deliberately minimal learning pipeline, are exercised on live, Zeek-monitored traffic(Zeek Project). The purpose is to validate dataset selection, not to engineer a production-grade IDS. We therefore asked a narrow question: with the learner family held constant, does the dataset with higher framework scores yield more stable detections across finance-relevant reconnaissance techniques when confronted with real network flows captured in Zeek `conn.log`? Zeek was chosen to provide transparent, per-flow ground truth and auditability via unique identifiers and a well-documented logging framework. Figure 5 summarizes the experimental workflow, with dataset selection and training at the top and Zeek integration for live detection at the bottom.

**Threat analysis and sector anchoring:** To ground the study in the financial sector, we first applied a targeted keyword lexicon ('*finance", banking", bank", credit", insurance", and financial"*') within the MITRE ATT&CK Enterprise database. Using our algorithm 1, we extracted finance-sector-relevant techniques that adversaries commonly employ against financial infrastructures. Table XVII presents six representative reconnaissance scenarios selected from this analysis. These scenarios represent the attack patterns most frequently employed against financial institutions and ensure both sector alignment and experimental tractability.

| Scenario | ATT&CK ID | Nmap options / rationale |
|---|---|---|
| TCP SYN port scan | T1595 (Active Scanning) | -sS; TCP SYN probe reconnaissance |
| Service/version detection | T1595.002 (Vulnerability Scanning) | -sV; bannering and service fingerprinting |
| Host discovery sweep | T1595.001 (Scanning IP Blocks); T1590 (Victim Network Information) | -sn; ping/ARP/ND sweeps |
| Stealth fragmented SYN | T1595 (evasive variant) | -sS -f -T2; fragmentation + lower timing |
| UDP port scan | T1595 (UDP) | -sU; protocol-diverse probing |
| Aggressive fingerprinting | T1592.002 (Victim Host Software); T1595.002 | -A; OS detection + scripts + versioning |

**Experimental setup and Method:** We trained two supervised flow-based detectors under the same minimal pipeline to see how dataset choice affects live behavior. One was trained on NSL-KDD and the other on CIC-UNSW-NB15 but both used Random Forest and the same flow features from Zeek's `conn.log`, without payloads, calibration, or threshold tuning as detailed in Table XVIII. A Zeek sensor captured all traffic and produced the `conn.log` records. Predictions were made per flow using a fixed classification probability threshold of 0.50. In practice, flows with a predicted attack probability of 0.50 or higher were labelled as attacks, while those below 0.50 were labelled as benign, and all outputs were linked to Zeek UIDs for traceability. Each scenario followed a repeatable cycle of baseline, attack window, and clean interval, as outlined in Table XIX to ensure reproducible ground truth establishment and consistent traffic capture conditions. Within these marked timeframes, we generated predictions, built confusion matrices, and summarized precision, recall, and F1 for attacks, measured false positives on benign traffic, and compared per-scenario recall to assess stability across reconnaissance variants.

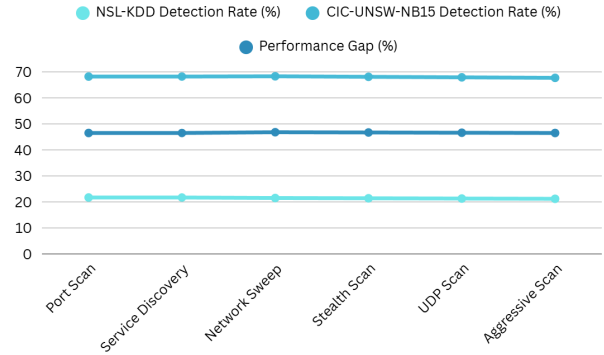

Fig. 6. Per-scenario detection rate and performance gap on live Zeek traffic

TABLE XVIII
Comparison of detectors trained on NSL-KDD and CIC-UNSW-NB15.

| Aspect | NSL-KDD-trained detector | CIC-UNSW-NB15-trained detector |
|---|---|---|
| Dataset era | Legacy, KDD'99 lineage | Contemporary, diverse lineage |
| Framework stance | Lower-rated dataset | Higher-rated dataset |
| Learner | Random Forest | |
| Features | Zeek `conn.log` flow fields | |
| Threshold | 0.50 (fixed) | |
| Audit linkage | Zeek UID join | |

TABLE XIX
Execution cycle and traffic capture in the Zeek-monitored testbed.

| Step | Description |
|---|---|
| 1 | Baseline: 30–60s of normal traffic (web, DNS, SSH). |
| 2 | Attack start: Begin 30s scan window. |
| 3 | Execute Nmap scenario (e.g., TCP SYN, UDP, service/version, aggressive). |
| 4 | Attack end: Stop at 30s mark. |
| 5 | Clean interval: 30–60s of normal traffic. |
| 6 | Ground truth: Mark timeframes in `conn.log` for UID correlation. |

**Results:** Figure 6 shows the detection rates for both CIC-UNSW-NB15 and NSL-KDD, and also the consistent performance gap.

Across six reconnaissance scenarios, the CIC-UNSW-NB15–trained detector achieved consistently higher recall and F1, while the NSL-KDD model showed higher precision (Table XX). CIC-UNSW-NB15 reached recall of 1.000 in every case with F1 between 0.831 and 0.835 and precision between 0.71 and 0.72. NSL-KDD recorded recall between 0.438 and 0.447, F1 between 0.609 and 0.618, and a precision of 1.000.

At the macro level, NSL-KDD averaged Precision = 1.000, Recall = 0.443, and F1 = 0.614, while CIC-UNSW-NB15 averaged Precision = 0.712, Recall = 1.000, and F1 = 0.832, yielding a +0.218 F1 advantage for CIC-UNSW-NB15. Considering all 232,905 flows across scenarios, CIC-UNSW-NB15 produced 158,526 positives (68.1%) versus 49,993 positives (21.5%) for NSL-KDD.

While our detection rates are stable and markedly higher for CIC-UNSW-NB15, they do not reach the 90% or near-perfect levels reported in prior work. Earlier work with CIC-family datasets confirms that performance can be greatly improved when complemented by engineering approaches such as calibration, oversampling, stacked embeddings, and PCA (Talukder et al., 2024, 2023; Kasongo and Sun, 2020; Harahsheh et al., 2024; Viegas et al., 2020). This body of work also indicates that optimisation is most effective after selecting a relevant, diverse dataset (Sarhan et al., 2021; Kenyon et al., 2020). Our results should therefore be read as a conservative baseline that can be improved through standard engineering.

**Interpretation with respect to the framework:** The outcomes align with the framework's finance-weighted predictions. Because both detectors used the same learner, the same Zeek `conn.log` feature family, and a fixed threshold, the observed performance gap is attributable to dataset choice rather than model design, reflecting the influence of dataset age, concept drift, recording environment, and split strategy on IDS evaluations (Ring et al., 2019). Each framework metrics maps to the observed behavior without repeating the numeric results of Table XX. Higher *ARS* and *TRS* for CIC-UNSW-NB15, driven by its coverage of contemporary finance-relevant scanning behaviors, are consistent with its uniform detection across scenarios. Stronger *TeRS* follows from closer alignment to Zeek's connection-level flow semantics, which were the sole feature source. *DQS* advantages, including clearer class definitions and more representative traffic, support stable detection across variants. *ECS* is not expressed in per-scenario rates but Zeek UID joins ensured that every prediction links to a specific flow in `conn.log`, providing auditability that supports compliance. Table XXI summarizes these mappings between finance expectations and observed signals.

TABLE XX
Per-scenario live detection performance for NSL-KDD- and CIC-UNSW-NB15-trained Random Forest models.

| Scenario | Total flows | NSL detections (n%) | NSL Precision | NSL Recall | NSL F1 | CIC detections (n%) | CIC Precision | CIC Recall | CIC F1 |
|---|---|---|---|---|---|---|---|---|---|
| Port scan | 38,410 | 8,330 (21.7%) | 1.000 | 0.447 | 0.618 | 26,205 (68.2%) | 0.711 | 1.000 | 0.831 |
| Stealth scan | 38,899 | 8,333 (21.4%) | 1.000 | 0.442 | 0.613 | 26,500 (68.1%) | 0.712 | 1.000 | 0.832 |
| Service discovery | 38,475 | 8,330 (21.7%) | 1.000 | 0.446 | 0.617 | 26,228 (68.2%) | 0.712 | 1.000 | 0.831 |
| Network sweep | 38,695 | 8,330 (21.5%) | 1.000 | 0.444 | 0.615 | 26,412 (68.3%) | 0.711 | 1.000 | 0.831 |
| Aggressive scan | 39,273 | 8,335 (21.2%) | 1.000 | 0.438 | 0.609 | 26,606 (67.7%) | 0.716 | 1.000 | 0.835 |
| UDP scan | 39,153 | 8,335 (21.3%) | 1.000 | 0.439 | 0.610 | 26,575 (67.9%) | 0.715 | 1.000 | 0.834 |

TABLE XXI
Mapping of framework metrics to finance expectations and observed outcomes. Higher framework scores for CIC-UNSW-NB15 anticipated its superior performance relative to NSL-KDD.

| Evaluation Metrics | Finance expectation | Observed signal | Evidence |
|---|---|---|---|
| ARS | Coverage of finance-relevant ATT&CK Recon (e.g., T1595, T1595.002) should improve detection | Stable detection across diverse scans | CIC recall = 1.000 for all six scenarios vs NSL ≈ 0.44; Table XX |
| TRS | Contemporary patterns and evasive variants should be represented | No degradation on stealth or aggressive probes | CIC sustained recall = 1.000 on fragmented SYN, UDP, and aggressive modes |
| TeRS | Features should align with Zeek `conn.log` and finance topologies | Robust flow-based detection without payloads | Both detectors used identical features; CIC generalized effectively |
| DQS | Balanced labels and quality features should support calibration | Precision–recall trade-off tunable post hoc | CIC precision = 0.712, FPR = 0.380 at 0.50 threshold; calibration can reduce FPR |
| ECS | Dataset must enable regulated-sector deployment | Governance and privacy suitability documented | Assessed at selection; not evidenced by per-scenario metrics. Each prediction is linked to a unique Zeek UID, enabling per-flow audit trails suitable for compliance review. |

## VIII. CONCLUSION & FUTURE WORKS

This paper has introduced a structured framework for evaluating the suitability of datasets in AI-based intrusion detection and prevention systems. By integrating quantitative measures, ARS, TRS, TeRS, ECS, and DQS, with the MITRE ATT&CK framework, the approach offers a more complete alternative to conventional evaluation methods. The application of this framework to benchmark datasets such as NSL-KDD, CICIDS2017, CICIDS2018, UNSW-NB15, and CICIoMT24 revealed notable differences in their capacity to represent real threat environments. The findings make clear that dataset selection should be tailored to sector-specific needs, with attention to technical, regulatory, and ethical factors. They also demonstrate that accuracy metrics alone are inadequate for assessing operational value; factors such as attack relevance, dataset currency, technical representativeness, and compliance are equally decisive. In future work, this project will actively collaborate with industry stakeholders and the MITRE ATT&CK initiative to scale the framework toward industry-level deployment and real-world adoption.

## IX. DECLARATION OF ORIGINALITY AND VERIFICATION

The authors confirm that this manuscript is their original work, has not been published elsewhere, and is not under consideration by any other journal or conference. All data, figures, and tables included herein are either the authors' own or have been cited appropriately. Each author listed has read and approved the final version of this manuscript, and there are no undisclosed conflicts of interest.

**Funding:** The authors received no specific grant from any funding agency in the public, commercial, or not-for-profit sectors.

**Conflict of Interest:** The authors declare that they have no conflict of interest.